%% file: ms.tex
\documentclass[]{aastex631}
\usepackage{CJK}
\usepackage{graphicx}
\usepackage{tabularx}
\usepackage{booktabs}
\usepackage{multirow}
\usepackage{makecell}
\usepackage{rotating}
\usepackage{subfigure}
\usepackage{gensymb}
\usepackage{ulem}

\shorttitle{IRAS 19312+1950 monitoring}
\graphicspath{{./}{figures/}}

\begin{document}
\begin{CJK*}{UTF8}{gkai}
\title{Considerations on the Origin of IRAS 19312+1950 Based on Long-Term Maser Observations}

\correspondingauthor{Jun-ichi Nakashima}
\email{junichin@mail.sysu.edu.cu, nakashima.junichi@gmail.com}

\author[0000-0001-7462-5968]{Huan-Xue Feng (冯焕雪)}
\affiliation{School of Physics and Astronomy, Sun Yat-sen University, Tang Jia Wan, Zhuhai, 519082, P. R. China}

\author[0000-0003-3324-9462]{Jun-ichi Nakashima(中岛淳一)}
\affiliation{School of Physics and Astronomy, Sun Yat-sen University, Tang Jia Wan, Zhuhai, 519082, P. R. China}
\affiliation{CSST Science Center for the Guangdong-Hong Kong-Macau Greater Bay Area, \\Sun Yat-Sen University, 2 Duxue Road, Zhuhai 519082, Guangdong Province, PR China}

\author[0000-0001-7102-6660]{D. Engels}
\affiliation{Hamburger Sternwarte, Universit\"{a}t Hamburg, Gojenbergsweg 112, D-21029 Hamburg, Germany}

\author[0000-0003-3483-6212]{S. Etoka}
\affiliation{Jodrell Bank Centre for Astrophysics, Department of Physics and Astronomy, University of Manchester, M13 9PL, UK}

\author[0000-0001-9825-7864]{Jaeheon Kim}
\affiliation{Korea Astronomy and Space Science Institute,776 Daedeok-daero, Yuseong-gu, Daejeon 34055, Republic of Korea}

\author[0000-0002-1086-7922]{Yong Zhang (张泳)}
\affiliation{School of Physics and Astronomy, Sun Yat-sen University, Tang Jia Wan, Zhuhai, 519082, P. R. China}
\affiliation{CSST Science Center for the Guangdong-Hong Kong-Macau Greater Bay Area, \\Sun Yat-Sen University, 2 Duxue Road, Zhuhai 519082, Guangdong Province, PR China}

\author[0000-0003-1015-2967]{Jia-Yong Xie (谢嘉泳)}
\affiliation{School of Physics and Astronomy, Sun Yat-sen University, Tang Jia Wan, Zhuhai, 519082, P. R. China}

\author[0000-0002-9829-8655]{Jian-Jie Qiu (邱建杰)}
\affiliation{School of Mathematics and Physics, Jinggangshan University, 28 Xueyuan Road, Qingyuan District, Ji’an 343009, Jiangxi Province, China}
\affiliation{School of Physics and Astronomy, Sun Yat-sen University, Tang Jia Wan, Zhuhai, 519082, P. R. China}
\affiliation{CSST Science Center for the Guangdong-Hong Kong-Macau Greater Bay Area, \\Sun Yat-Sen University, 2 Duxue Road, Zhuhai 519082, Guangdong Province, PR China}

\begin{abstract}

IRAS source 19312+1950 (hereafter I19312) is an infrared point source with maser emissions of SiO, H$_2$O, and OH molecules. Although initial observations suggested that I19312 might be an evolved star, its characteristics are not fully consistent with this classification. This study aims to further investigate the nature of I19312 by conducting long-term monitoring of its maser emissions and comparing the results with other known astrophysical objects. We conducted long-term monitoring of SiO, H$_2$O, and OH maser emissions using single-dish radio telescopes. The results were then compared with historical maser data and the characteristics of similar objects to infer the possible origin of I19312. The SiO maser emissions from I19312 were detected over a wide velocity range and exhibited significant time variability. The OH maser lines suggest characteristics of an evolved star, while the H$_2$O maser lines indicate molecular outflows. These features suggest that I19312 could be a candidate for a Water Fountain (WF) star, though there are inconsistencies, such as the large molecular gas mass, that challenge this hypothesis. The possibility of I19312 being a Red Nova Remnant (RNR) is also considered, but this remains speculative due to the lack of direct evidence. The evolutionary stage of I19312 remains unclear, but it shares multiple characteristics with both evolved stars with peculiar properties and RNRs. Further long-term monitoring and high-resolution interferometric observations are required to better constrain the nature of this object.

\end{abstract}

\keywords{stars: atmospheres --- 
stars: chemically peculiar --- 
(stars:) circumstellar matter --- 
stars: individual (IRAS 19312+1950)}

\section{Introduction} \label{sec:intro}
Maser emissions of SiO, H$_2$O, and OH molecules have been detected in the IRAS source 19312+1950  \citep[hereafter referred to as I19312,][]{2000PASJ...52L..43N,2011ApJ...728...76N}. IRAS sources where maser emissions of SiO, H$_2$O, and OH molecules are detected (especially when SiO maser emissions are present) are often evolved stars with active mass loss, such as asymptotic giant branch (AGB) stars or red supergiants (RSGs). Indeed, various observations have been conducted on I19312, and many of the results align well with the characteristics of an evolved star with a cool circumstellar envelope.

However, not all observed properties match those of an evolved star. For instance, mapping observations in the CO radio line using the Nobeyama 45~m telescope \citep{2016ApJ...825...16N} revealed that the total mass of molecular gas ranges between 225 and 478 M$_\odot$, which is considerably large. Such a massive accumulation of molecular gas is difficult to explain with a single AGB star or RSG. However, the same observation showed that the CO gas is isolated near I19312, suggesting it is different in nature from a young stellar object (YSO) within a star-forming region. The point-symmetric nebula observed at the near-infrared bands also exists in isolation \citep{2004PASJ...56.1083D,2007A&A...470..957M}, and the shape and isolation of the nebula seems to match well the expected characteristics of an evolved star. Methanol lines have been detected from I19312 \citep{2004PASJ...56.1083D,2015PASJ...67...95N}. While methanol is often detected in YSOs, it has been also detected from an evolved star as well \citep{2017A&A...603L...2O}, indicating that methanol is not a definitive factor for distinguishing between an evolved star and a YSO.

The distance to I19312 obtained from the annual parallax measurement using VLBI ($3.80^{+0.83}_{-0.58}$ kpc) imposes significant constraints on the YSO scenario. Considering this distance, the total infrared flux of I19312 excludes the possibility of it being a low-mass YSO \citep{2016ApJ...825...16N}. Furthermore, all YSOs detected so far with SiO masers are massive YSOs \citep{1984AJ.....89.1833B,2009ApJ...691..332Z,2016ApJ...826..157C}, and massive YSOs generally have associated HII regions. If an HII region were present, radio continuum emission from free-free radiation or the Br${\gamma}$ line should be detected, but these features are not observed in I19312 \citep[see, e.g.,][]{2011ApJ...728...76N}. Additionally, in massive YSOs, the OH maser emission at 1665 MHz and 1667 MHz (i.e., main lines) is typically brighter than the 1612 MHz line \citep[i.e., satellite lines,][]{1996A&ARv...7...97H}. However, in I19312, the 1612 MHz line is brighter than the main lines \citep{2011ApJ...728...76N}. Moreover, radio observations of molecular lines suggest a relatively high abundance of $^{13}$C compared to $^{12}$C \citep{2004PASJ...56..193N,2004PASJ...56.1083D,2023A&A...669A.121Q}. Considering these observational characteristics comprehensively, I19312 tends to exhibit properties more similar to an evolved star than a YSO. However, even though the possibility of it being a YSO is low, there are still some observational properties that cannot be easily explained by a typical evolved star. It has been proposed that I19312 may be a rare case of an evolved star existing in a star-forming region with the same line-of-sight velocity \citep{2016ApJ...825...16N}, but from a probabilistic standpoint, this possibility does not seem very likely.

Another interpretation, different from evolved stars and YSOs, is that I19312 might be a red nova remnant \citep[hereafter RNR; see, e.g.,][]{2018A&A...617A.129K}. A red nova is an explosive event that occurs when two main-sequence stars merge \citep[note: it has also been suggested that a merger involving evolved stars also may trigger a red nova event. See, e.g.,][and references therein]{2018A&A...617A.129K}, and after the red nova merger event, a cool molecular envelope is known to form (called a red nova remnant, RNR). Additionally, SiO masers have been detected from a RNR \citep{2005PASJ...57L..25D,2007ASPC..363...87C}. To date, five RNRs have been confirmed: CK Vul \citep{2015Natur.520..322K}, V1309 Sco \citep{2011A&A...528A.114T,2016A&A...592A.134T}, OGLE-2002-BLG-360 \citep{2013A&A...555A..16T}, V838 Mon \citep{2006A&A...451..223T,2021A&A...655A..32K}, and V4332 Sgr \citep{2015A&A...578A..75T}. In the case of these objects, the fact that they are RNRs is well supported by the direct observation of explosive events caused by stellar mergers. On the other hand, there is no definitive evidence that I19312 experienced an explosive event in the past \citep{2023A&A...669A.121Q}. The infrared characteristics of RNRs and evolved stars are very similar, and in the absence of a record of a red nova event, it is difficult to distinguish RNRs from evolved stars based on infrared characteristics alone. One suggests that a significant number of RNRs could be contaminated with evolved stars \citep{2020A&A...644A..59K}.

In this study, we conducted long-term monitoring observations in the SiO, H$_2$O, and OH maser lines towards I19312 using single-dish radio telescopes. We also combined the present observation results with all previous maser observations of I19312 and compared them with the maser characteristics of astrophysical objects with similarities to I19312. At the time when SiO masers were detected from I19312 in 2000, SiO masers were primarily known to be detected from evolved stars and a small number of YSOs. However, over the more than 20 years since then, SiO maser emissions have been detected from several new types of objects, including RNRs. In this paper, we discuss the origin of I19312 by comparing its maser characteristics with those of these new types of SiO maser sources.

\section{Observations \& Data Reduction} \label{sec:observations and data reduction}

I19312 was observed from February to September 2018 using the Nan\c{c}ay Radio Telescope (NRT) to monitor the four OH maser lines (1612\,MHz, 1665\,MHz, 1667\,MHz, 1720\,MHz). Additionally, observations were conducted from January 2018 to May 2020 using the 21-m single-dish telescopes of the Korean VLBI Network (KVN). The SiO $v = 1$, $J = 1$--0, 2--1, 3--2, and $v = 2$, $J = 1$--0 (rest frequencies are 43.122\,GHz, 86.243\,GHz, 129.363\,GHz, and 42.821\,GHz, respectively), and H$_2$O $6_{1,6} - 5_{2,3}$ (rest frequency is 22.235\,GHz) maser lines were observed using the KVN single-dish telescope. The position of I19312 used for the maser observations and the observed maser lines are summarized in Table~\ref {tab:Target and Observed Lines}. Table~\ref{tab: obs date} lists the observation dates, where ``TD'' denotes the total number of days counted from January 3, 2018, when the present monitoring observation began.

The NRT is a transit telescope, with a half-power beam width at 1.6\,GHz of 3.5$'$ in right ascension (RA) by 19$'$ in declination (Dec). The system temperature was typically 35\,K, and observations were carried out in frequency switching mode. The 8192-channel digital auto-correlator was divided into 8 banks of 1024 channels each, enabling the simultaneous observation of the 4 transitions in both left- and right-handed circular polarizations (LHC and RHC). A bandwidth of 0.78\,MHz was employed for each bank. The spectra were Hanning-smoothed during reduction, resulting in a velocity resolution of $\sim$ 0.13--0.14\,km\,s$^{-1}$ for the four OH lines. The ratio of flux to antenna temperature was 1.4\,K\,Jy$^{-1}$ at 0$^{\circ}$ declination. The instrumental polarization is regularly checked by the NRT staff, and the uncertainty in the absolute flux density is $\sim$10$\%$. The temporal variations in the four OH lines in total flux density (Stokes\,$I$) and circularly polarized flux density (Stokes\,$V$) can be determined from the flux densities $F$ of both circular polarizations in each frequency channel of the spectrum analyzer, as described in \citet{2010ARep...54..599S}: Stokes\,$I$ was calculated as $F(\text{RHC})+F(\text{LHC})$.

The KVN receiver optics system was configured for simultaneous observations in four bands, namely, the H$_2$O\,22\,GHz and SiO\,43, 86, and 129\,GHz bands (see, KVN Status Report\footnote{\url{https://radio.kasi.re.kr/status_report.php}}). The average half-power beam widths (HPBW)/aperture efficiencies for the four bands were 123$''$/0.58 (22\,GHz), 62$''$/0.61 (43\,GHz), 32$''$/0.50 (86\,GHz), and 23$''$/0.35 (129\,GHz). A position-switching mode was adopted, with pointing and focus checked approximately every 2 -- 3 hours by observing nearby strong SiO maser sources. A high-electron-mobility transistor (HEMT) receiver was used for the 22, 43, and 86\,GHz bands, while a superconductor-insulator-superconductor (SIS) receiver was used for the 129\,GHz band. Both receivers could simultaneously receive both LCP and RCP circular polarization \citep{2013PASP..125..539H}, although only the LCP was observed in this study. The system noise temperature of 22, 43, 86, and 129\,GHz bands was maintained within the range of 80--250\,K (22\,GHz), 100--280\,K (43\,GHz), 200--400\,K (86\,GHz), and 140--630\,K (129\,GHz), depending on weather conditions and elevation of the target source. Digital spectrometers with a 32\,MHz bandwidth were used for the 22\,GHz and 43\,GHz bands, and with a 64\,MHz bandwidth for the 86\,GHz and 129\,GHz bands. The velocity coverage and resolutions were 440 and 0.11\,km\,s$^{-1}$ (22\,GHz),  222 and 0.05\,km\,s$^{-1}$ (43\,GHz),  222 and 0.05\,km\,s$^{-1}$ (86\,GHz), and 148 and 0.036\,km\,s$^{-1}$ (129\,GHz). To improve the SNR, the spectra were Hanning-smoothed to a velocity resolution of $\sim$ 0.43\,km\,s$^{-1}$. The intensity was calibrated using the chopper wheel method to correct for atmospheric attenuation and variations in antenna gain, followed by sky dipping curve analysis to correct the atmospheric opacity above the antenna, yielding the antenna temperature $T_{A}^{*}$. The conversion factors from K to Jy are 13.8\,Jy\,K$^{-1}$ (22\,GHz), 13.1\,Jy\,K$^{-1}$ (43\,GHz), 15.9\,Jy\,K$^{-1}$ (86\,GHz), and 22.8\,Jy\,K$^{-1}$ (129\,GHz).

\section{Results} \label{sec: results}

\subsection{Observational Results of OH Maser Lines and Comparison with Past Observations} \label{sec: oh}
The four OH maser lines (1612 MHz, 1665 MHz, 1667 MHz, and 1720 MHz lines) were observed eight times over 241 days (see Table~\ref{tab: obs date}). The observation results of the OH 1612, 1665, and 1667 MHz lines are summarized in Table~\ref{tab: OH1612_65_67MHz parameters}. Table~\ref{tab: OH1612_65_67MHz parameters} includes the observation date, the minimum and maximum radial velocities ($V_{\rm min}$ and $V_{\rm max}$) of the channels where the line emission were detected, the velocity range ($V_{\rm range}$) of the line emission, the peak flux density ($F_{\rm peak}$), the line-of-sight velocity at the flux density peak ($V_{\rm peak}$), the integrated flux density ($F_{\rm int}$), the rms noise level (rms), and the signal-to-noise ratio (SNR) at the flux density peak of each line. $V_{\rm min}$ and $V_{\rm max}$ are the values obtained from the range of channels where the detected signal exceeds 5$\sigma$. However, in the visual inspection of the spectra, some features were clearly detected even though their SNR slightly fell below 5$\sigma$. These cases are marked with an asterisk ``*'' in the table to distinguish them from values obtained from channels above 5$\sigma$. The 1720 MHz line was not detected, so only the rms noise is listed in Table~\ref{tab: OH1720_MHz parameters}.

\begin{figure}
    \centering
    \includegraphics[width=0.6\textwidth]{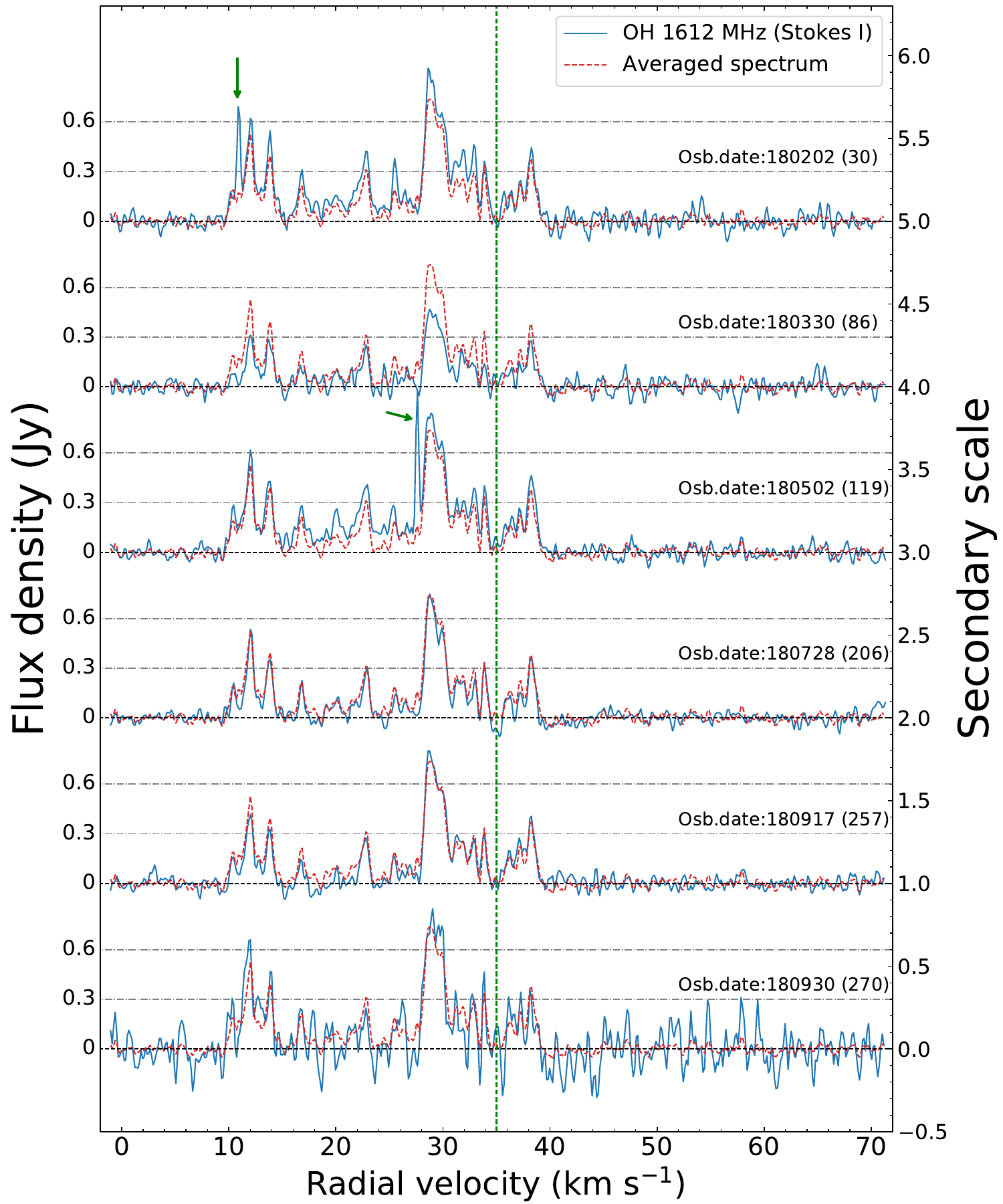}
    \caption{Spectra of the OH\,1612\,MHz maser line of I19312 in Stokes\,$I$ (blue line) from February 2, 2018 to September 30, 2018. The red broken line is the average of the spectra for all observation dates. The numbers in parentheses are the total number of days counted from January 3, 2018, the start date of the present monitoring observation. The green dotted vertical line represents the systemic velocity ($\sim$\,35\,km\,s$^{-1}$) of the broad component suggested by \citet{2005ApJ...633..282N} based on their BIMA observations in the CO lines. The green arrows indicate emission peaks that show a rapid time variation (see text). To display the line flux densities, individual scales for each spectrum are provided on the left side in units of Jy. Additionally, a common secondary scale is shown on the right side to facilitate reading the absolute flux density values. Since the spectra from adjacent observation dates (July 28-29, 2018, and September 17-18, 2018) are nearly identical, we present the averaged spectrum as the representative spectrum here. The spectrum from July 28, 2018, is the average of the spectra obtained on July 28 and 29, 2018, while the spectrum from September 17, 2018, is the average of the spectra from September 17 and 18. }
    \label{fig: OH1612MHz}		
\end{figure}

\begin{figure}
    \centering
    \includegraphics[width=0.6\textwidth]{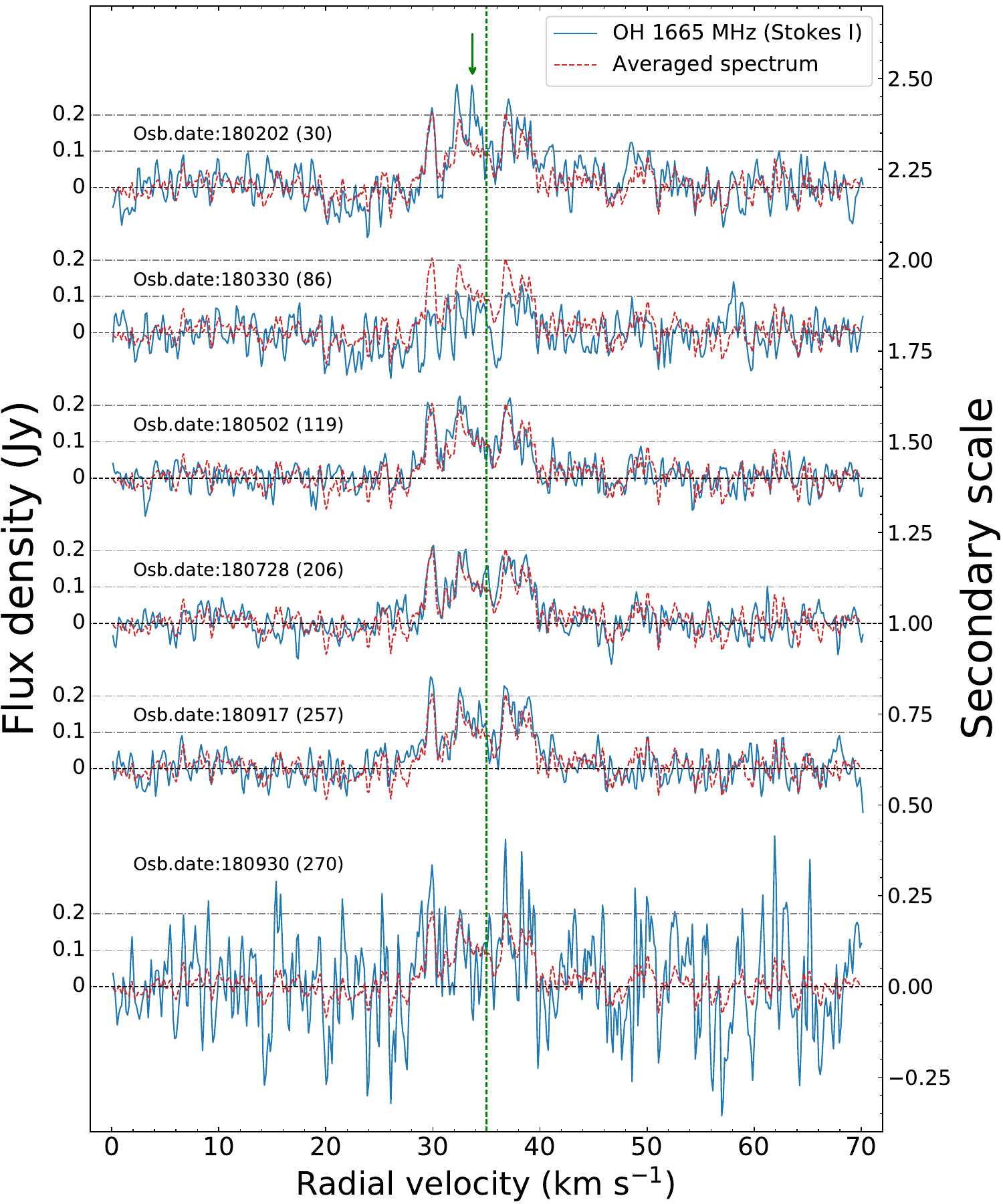}
    \caption{Spectra of the OH 1665 MHz maser line of I19312 in Stokes $I$ from February 2, 2018, to September 30, 2018. The notations are the same as in Figure~\ref{fig: OH1612MHz}. Note that features caused by RFI, which could not be completely removed during the reduction process, remain in the velocity range above 55 km~s$^{-1}$. These RFI-induced features are typically detected in only a single channel and are observed only on specific observation days.}
    \label{fig: OH1665MHz}			
\end{figure}

\begin{figure}
    \centering
    \includegraphics[width=0.6\textwidth]{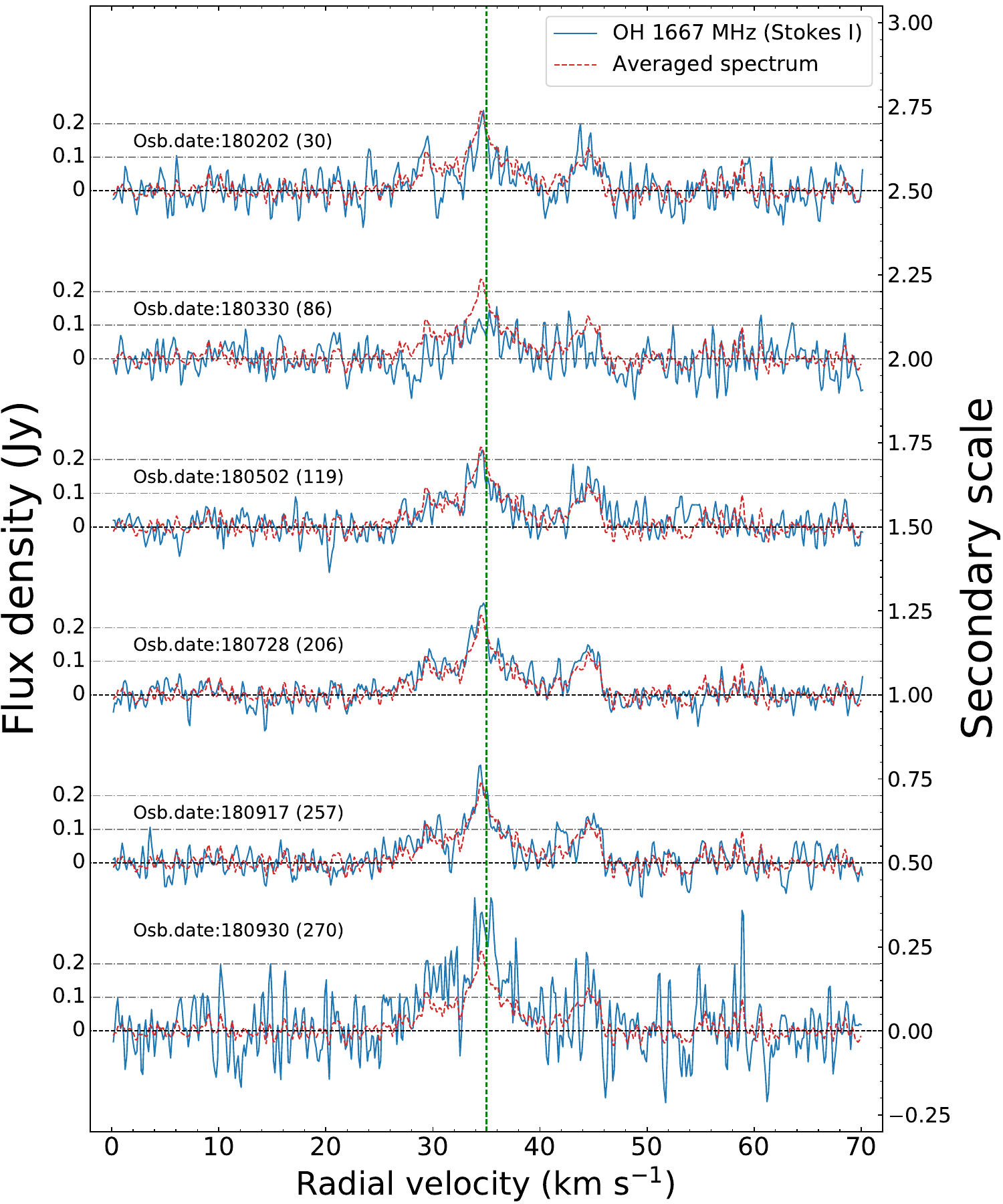}
    \caption{Spectra of the OH\,1665\,MHz maser line of I19312 in Stokes\,$I$ from February 2, 2018 to September 30, 2018. Notations are the same as in Figure~\ref{fig: OH1612MHz}. Note that features caused by RFI, which could not be completely removed during the reduction process, remain in the velocity range above 55 km~s$^{-1}$. These RFI-induced features are typically detected in only a single channel and are observed only on specific observation days.}
    \label{fig: OH1667MHz}				
\end{figure}

Figures~\ref{fig: OH1612MHz}, \ref{fig: OH1665MHz}, and \ref{fig: OH1667MHz} show the spectra of the OH 1612, 1665, and 1667 MHz lines, respectively. The OH 1612 MHz line (satellite line) and the OH 1665 MHz and 1667 MHz lines (main lines) were clearly detected in these  observations (except for one day with very poor weather conditions). The satellite line was found to be stronger than the main lines in both peak flux density and velocity-integrated flux density. The line intensity ratio between the satellite line and the main lines is consistent with past observations \citep{2011ApJ...728...76N,2014ApJ...794...81Y}.

\begin{figure}
    \centering
    \includegraphics[width=0.6\textwidth]{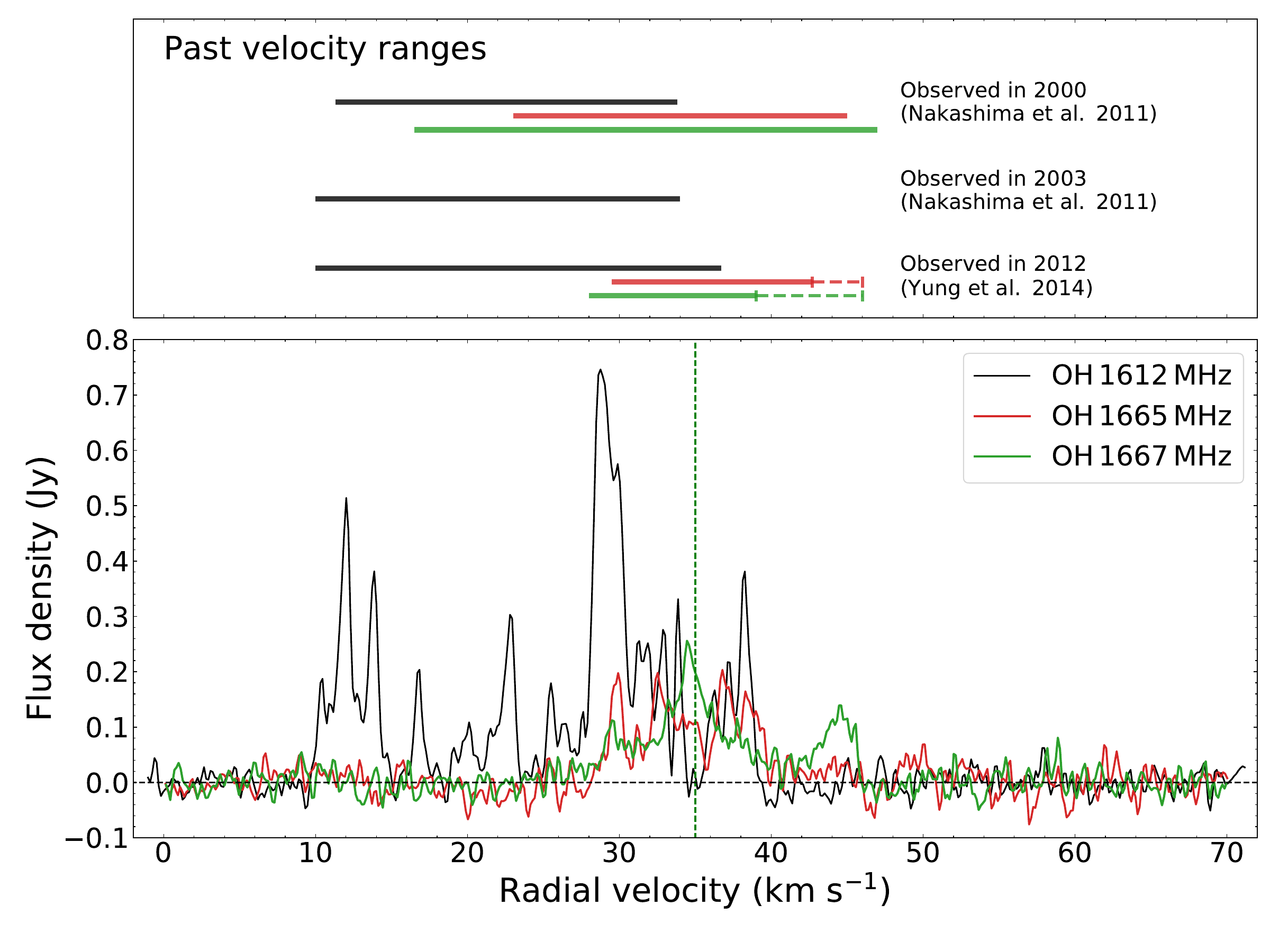}
    \caption{Upper panel: velocity ranges of previously detected OH maser lines. The colors of the horizontal lines correspond to the legend in the lower panel. The dashed parts of the horizontal lines represent the visually corrected velocity range (see main text). Bottom panel: comparison of the averaged OH maser lines (Stokes\,$I$) obtained in the present observations.}  
    \label{fig: OH-comp}			
\end{figure}

The average spectra calculated from the spectra of all observation days are shown in the lower panel of Figure~\ref{fig: OH-comp}. It is evident that the velocity range in which emission is detected  varies from one transition to the other. The velocity range of the OH 1612 MHz line is about 10 km~s$^{-1}$ to 40 km~s$^{-1}$, while that of the OH 1665 MHz  line is about 26 km~s$^{-1}$ to 40 km~s$^{-1}$ and that of the OH 1667 MHz  line is  about 26 km~s$^{-1}$ to 46 km~s$^{-1}$. A weak feature of the 1665 MHz line may be detected at about 50 km~s$^{-1}$ (see Figures~\ref{fig: OH1665MHz}). Other weak features of the main lines seem to exist in the velocity range above 55 km~s$^{-1}$, but these are confirmed to be Radio Frequency Interference (RFI) that could not be removed during the reduction process (note: RFI typically appears as narrow features of about one channel and is detected only on a single observation day).

\begin{figure}
    \centering
    \includegraphics[width=0.7\textwidth]{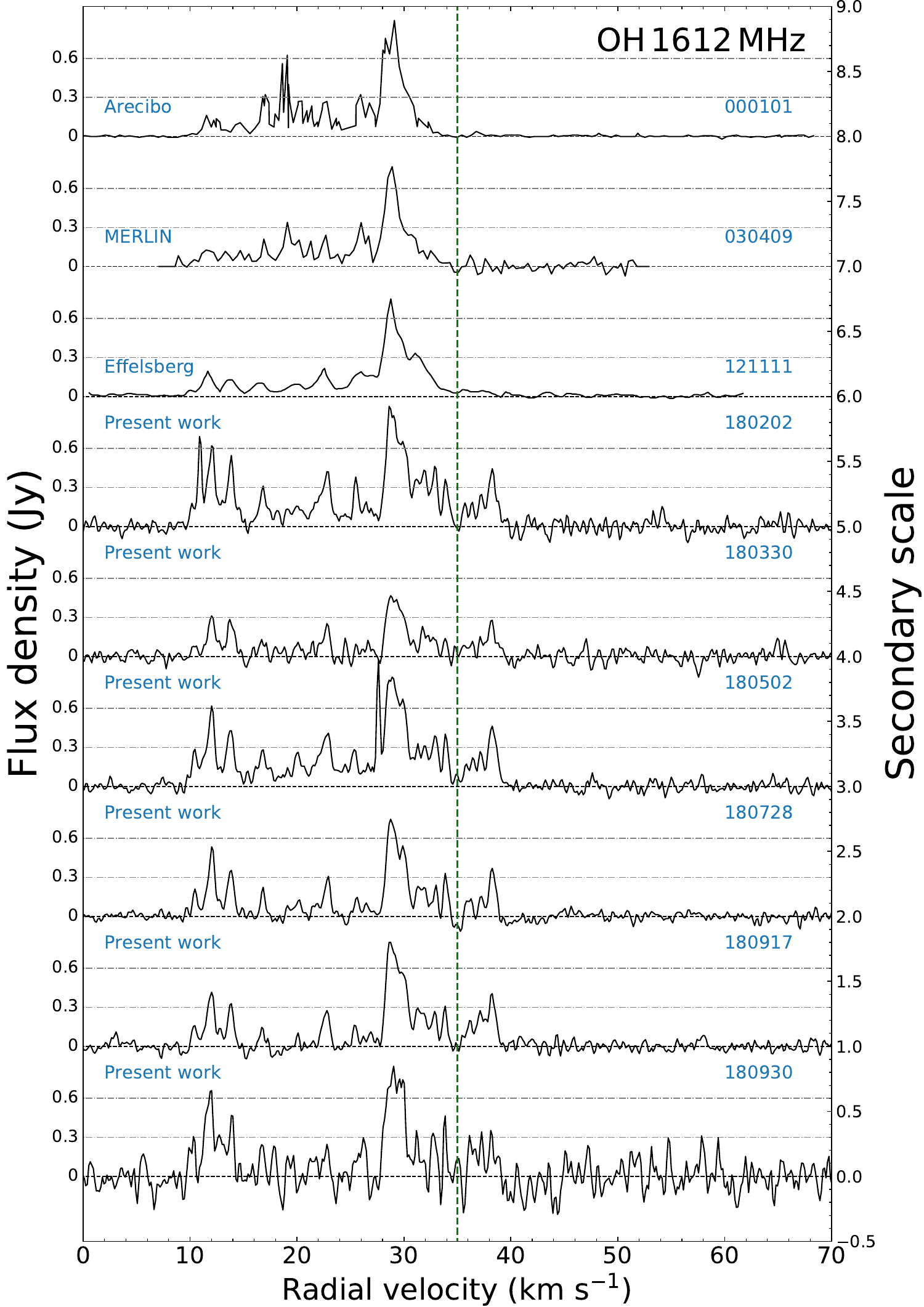}
    \caption{Time evolution of the OH\,1612\,MHz maser spectra of I19312 over the last 20 years. Arecibo and MERLIN data are taken from \citet{2011ApJ...728...76N}, and Effelsberg data are taken from \citet{2014ApJ...794...81Y}. The date of observation (YYMMDD) is given for each spectral line. The green dotted vertical line represents the systemic velocity of I19312 ($\sim$\,35\,km\,s$^{-1}$). To display the line flux densities, individual scales for each spectrum are provided on the left side in units of Jy. Additionally, a common secondary scale is shown on the right side to facilitate reading the absolute flux density values.}
    \label{fig: OH1612MHz-past-spectrum}				
\end{figure}

\begin{figure}
    \centering
    \includegraphics[width=0.7\textwidth]{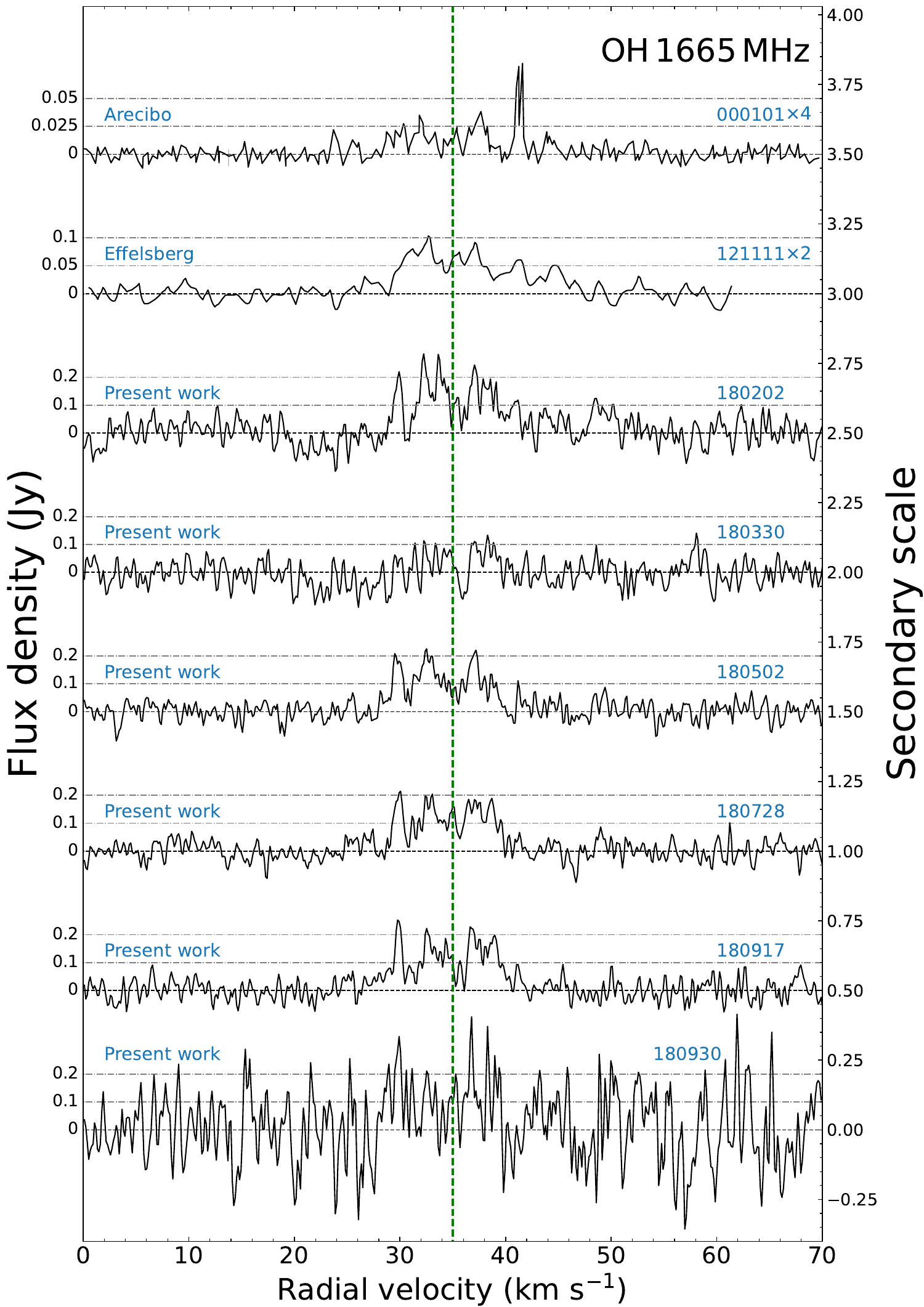}
    \caption{Time evolution of the OH\,1665\,MHz maser spectra of I19312 over the last 20 years. Arecibo data are taken from \citet{2011ApJ...728...76N}, and Effelsberg data are taken from \citet{2014ApJ...794...81Y}.  The date of observation (YYMMDD) is given for each spectral line. The green dotted vertical line represents the systemic velocity of I19312 ($\sim$\,35\,km\,s$^{-1}$). Past spectra are magnified along the vertical axis to view line profiles. Magnification is given to the right of the observation date. To display the line flux densities, individual scales for each spectrum are provided on the left side in units of Jy. Additionally, a common secondary scale is shown on the right side to facilitate reading the absolute flux density values.}
    \label{fig: OH1665MHz-past-spectrum}				
\end{figure}

\begin{figure}
    \centering
    \includegraphics[width=0.7\textwidth]{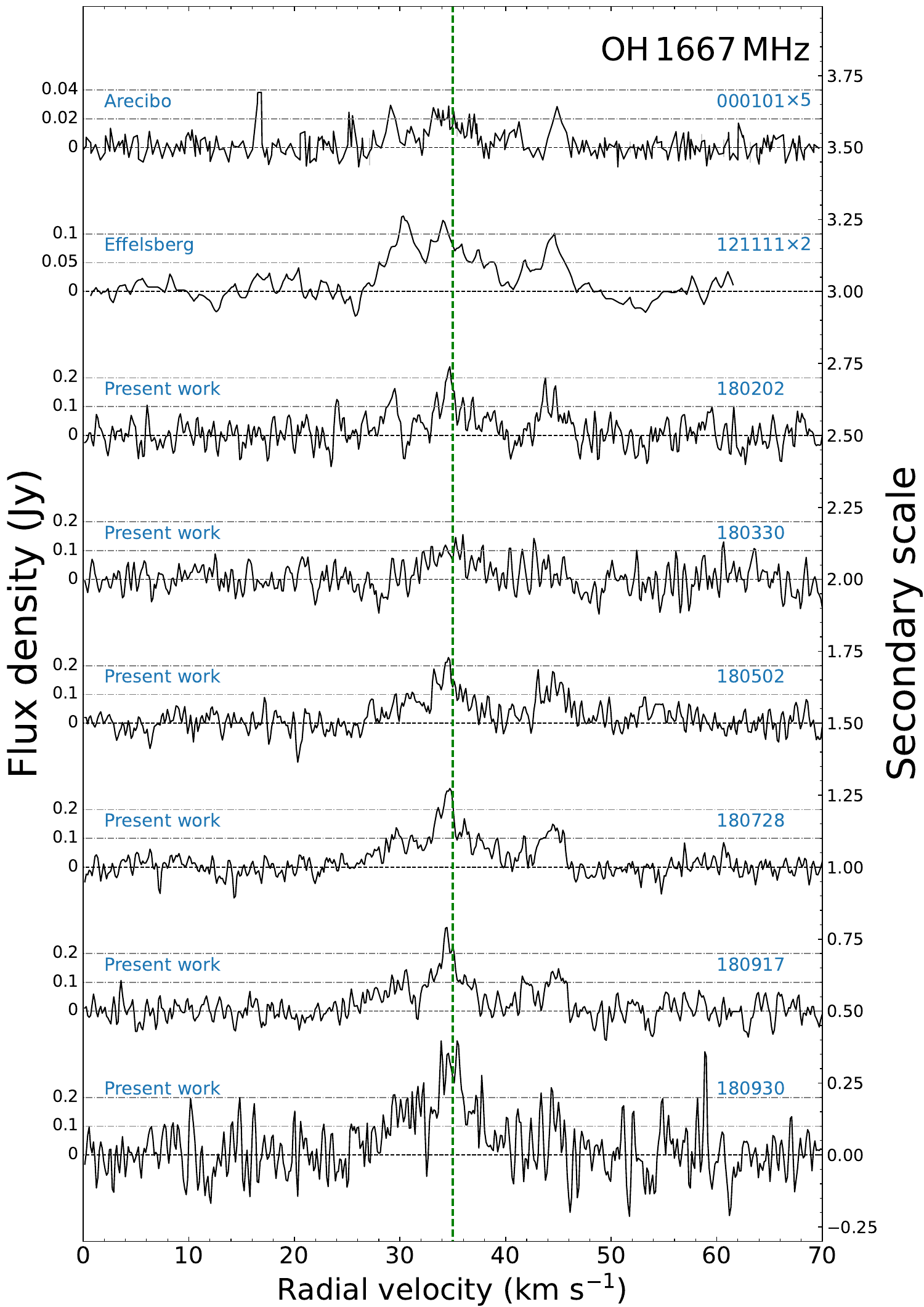}
    \caption{Time evolution of the OH\,1667\,MHz maser spectra of I19312 over the last 20 years. Arecibo data are taken from \citet{2011ApJ...728...76N}, and Effelsberg data are taken from \citet{2014ApJ...794...81Y}. The date of observation (YYMMDD) is given for each spectral line. The green dotted vertical line represents the systemic velocity ($\sim$\,35\,km\,s$^{-1}$). Past spectra are magnified along the vertical axis to view line profiles. Magnification is given to the right of the observation date. To display the line flux densities, individual scales for each spectrum are provided on the left side in units of Jy. Additionally, a common secondary scale is shown on the right side to facilitate reading the absolute flux density values.}
    \label{fig: OH1667MHz-past-spectrum}				
\end{figure}

Before the present observation, the OH maser lines at 1612 MHz, 1665 MHz, and 1667 MHz were observed in 2000, 2003, and 2012 using the Arecibo radio telescope, MERLIN, and the Effelsberg 100~m telescope \citep{2011ApJ...728...76N, 2014ApJ...794...81Y}. The velocity ranges of each OH maser line obtained from these observations are displayed in the upper panel of Figure~\ref{fig: OH-comp} for comparison. Additionally, to confirm changes in the line profiles, the spectra from past observations and the current observation are arranged chronologically in Figures~\ref{fig: OH1612MHz-past-spectrum}, \ref{fig: OH1665MHz-past-spectrum}, and \ref{fig: OH1667MHz-past-spectrum}. In the Arecibo observation, the OH 1612 MHz line was detected in the velocity range of approximately 11 km~s$^{-1}$ to 34 km~s$^{-1}$. The OH 1665 MHz and 1667 MHz lines were detected in the velocity ranges of approximately 23 km~s$^{-1}$ to 45 km~s$^{-1}$ and 17 km~s$^{-1}$ to 47 km~s$^{-1}$, respectively. In the Effelsberg 100~m telescope observation, the OH 1612 MHz line was detected in the velocity range of 10 km~s$^{-1}$ to 37 km~s$^{-1}$, and the OH 1665 MHz and 1667 MHz lines were detected in the velocity ranges of 30 km~s$^{-1}$ to 46 km~s$^{-1}$ and 28 km~s$^{-1}$ to 46 km~s$^{-1}$, respectively (note: careful inspection of the spectra published in the past paper reveals that weak features appear to be detected outside the velocity range reported by the authors. These are indicated by dashed lines in Figure~\ref{fig: OH-comp}). Comparing the spectra obtained from past observations with those from the present observation, it is evident that a new peak has appeared around 38 km~s$^{-1}$ in the spectrum of the 1612 MHz line. On the other hand, there was no significant change in the velocity range in which the 1665 MHz and 1667 MHz lines were detected. Summarizing the overall trend of OH maser lines over the past 20 years, the satellite line (1612 MHz line) is detected in a relatively low velocity range (blueshifted with respect to the systemic velocity), while the main lines (1665 MHz and 1667 MHz lines) are detected in a relatively high velocity range (redshifted with respect to the systemic velocity).

Close inspection of Figures~\ref{fig: OH1612MHz}, \ref{fig: OH1665MHz}, and \ref{fig: OH1667MHz} reveals that the intensities of many peaks change synchronously. For example, in the spectrum of the OH 1612 MHz line on February 2, 2018, almost all peaks are above average intensity, while in the spectrum on May 2, 2018, many peaks are below average intensity. Similar trends are observed in the spectra of the 1665 MHz and 1667 MHz lines. However, some peaks exhibit exceptional time variability. For instance, in the OH 1612 MHz line, the peaks around 27 km~s$^{-1}$ in the May 2, 2018 spectrum and around 11 km~s$^{-1}$ in the February 2, 2018 spectrum (see peaks marked with green arrows in Figure~\ref{fig: OH1612MHz}) were detected only on specific observation days and disappeared within a short time scale of 1-2 months. Similar situations are observed in the OH 1665 MHz line (see Figure~\ref{fig: OH1665MHz}). As we have sufficiently confirmed that there is no RFI in the velocity range where these peaks are detected, it is reasonable to consider the detected signals to be of astronomical origin. Given their relatively fast time variability, they are presumably unsaturated maser emissions.

\subsection{Observational Results of H$_2$O Maser Lines and Comparison with Past Observations} \label{sec: h2o}

\begin{figure}
    \centering
    \includegraphics[width=0.6\textwidth]{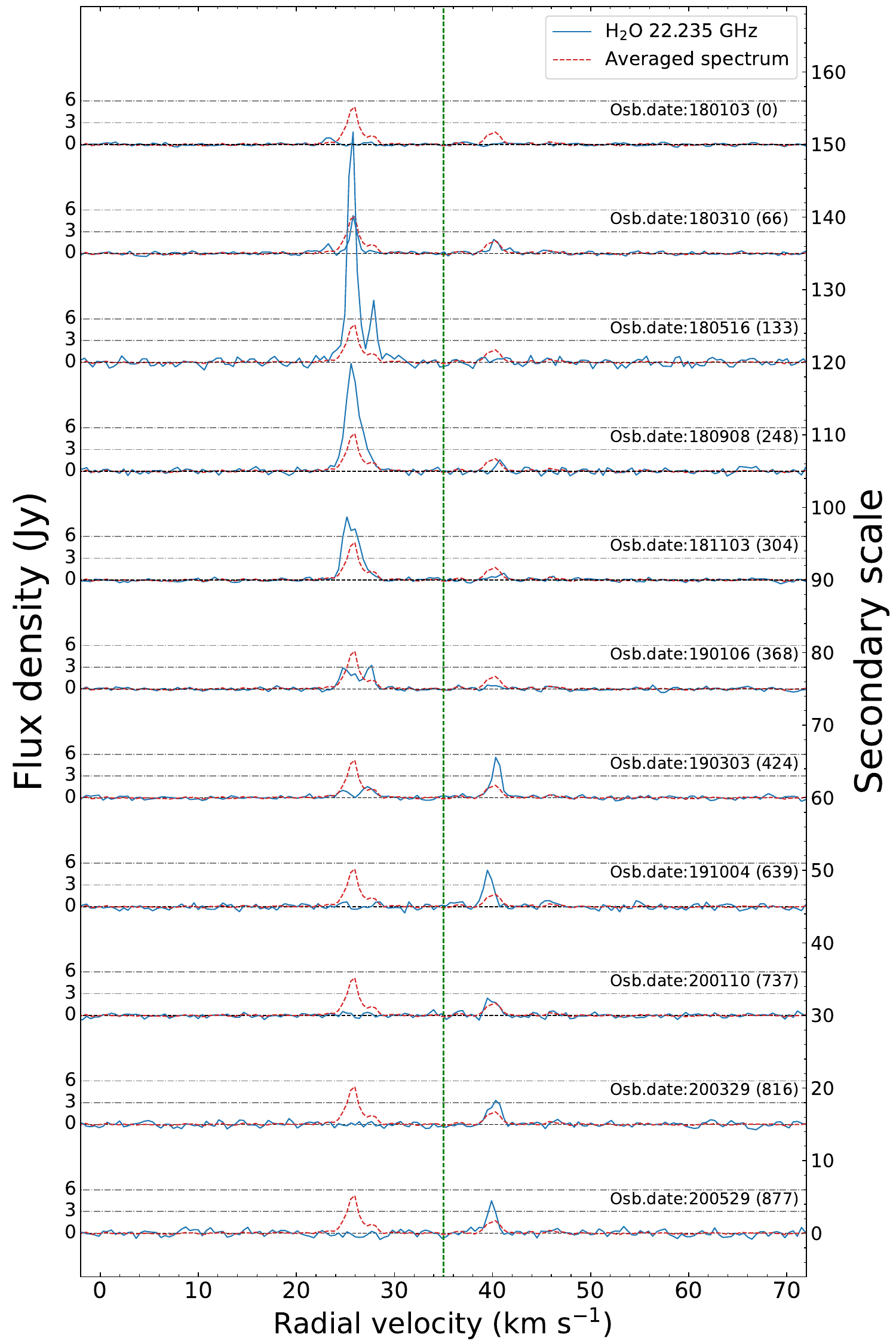}
    \caption{Spectra of the H$_{2}$O\,22.235\,GHz line of I19312 from 2018 January to 2020 May. Notations are the same as in Figure~\ref{fig: OH1612MHz}.}
    \label{fig: H2O22.2GHz}					
\end{figure}

The 22.235 GHz H$_2$O maser line and the four SiO masers lines listed in Table~\ref{tab:Target and Observed Lines} were simultaneously observed with the 21-m single-dish telescope of KVN. The observations were conducted 11 times over 817 days (see Table~\ref{tab: obs date}, and the observation results of the SiO masers are summarized in Section~\ref{sec: sio}). The H$_2$O spectra are shown in Figure~\ref{fig: H2O22.2GHz}, and the line parameters are summarized in Table~\ref{tab: H2O_blue_red parameters}. The meanings of the parameters listed in Table~\ref{tab: H2O_blue_red parameters} are essentially the same as those in Table~\ref{tab: OH1612_65_67MHz parameters}. However, as evident from Figure~\ref{fig: H2O22.2GHz}, the H$_2$O maser line exhibits a double-peak line profile, so in addition to the parameters for the entire line, the parameters for the redshifted and blueshifted components are listed separately.

From Figure~\ref{fig: H2O22.2GHz}, it can be seen that the H$_2$O maser line is detected in the range of about 22 km~s$^{-1}$ to 42 km~s$^{-1}$ (the blueshifted component is from about 22 km~s$^{-1}$ to 28 km~s$^{-1}$, and the redshifted component is from about 38 km~s$^{-1}$ to 42 km~s$^{-1}$). The observations cover a velocity range from $-180$ km~s$^{-1}$ to 250 km~s$^{-1}$, but outside the range shown in Figure~\ref{fig: H2O22.2GHz}, no other features were found except for a weak signal detected at about 78 km~s$^{-1}$ on March 3, 2019 (corresponding to 0.049 Jy at a peak of 3.2$\sigma$). The spectra clearly show that both the redshifted and blueshifted components exhibit significant time variability. Until January 2019, the intensity of the blueshifted component remained strong, but after that, the intensities switched, and the redshifted component became stronger. The blueshifted component shows a strong peak at 26 km~s$^{-1}$, with relatively weaker peaks at 25 km~s$^{-1}$ and 28 km~s$^{-1}$. The blueshifted component at 26 km~s$^{-1}$ began increasing in intensity around March 2018, reaching its peak in mid-May. Afterward, the intensity gradually decreased until it completely disappeared by March 2019. The redshifted component shows a single strong peak around 40 km~s$^{-1}$, without any additional weaker peaks appearing during the observation period, unlike the blueshifted component. 

\begin{figure}
    \centering
    \renewcommand{\thefigure}{9a}
    \includegraphics[width=0.7\textwidth]{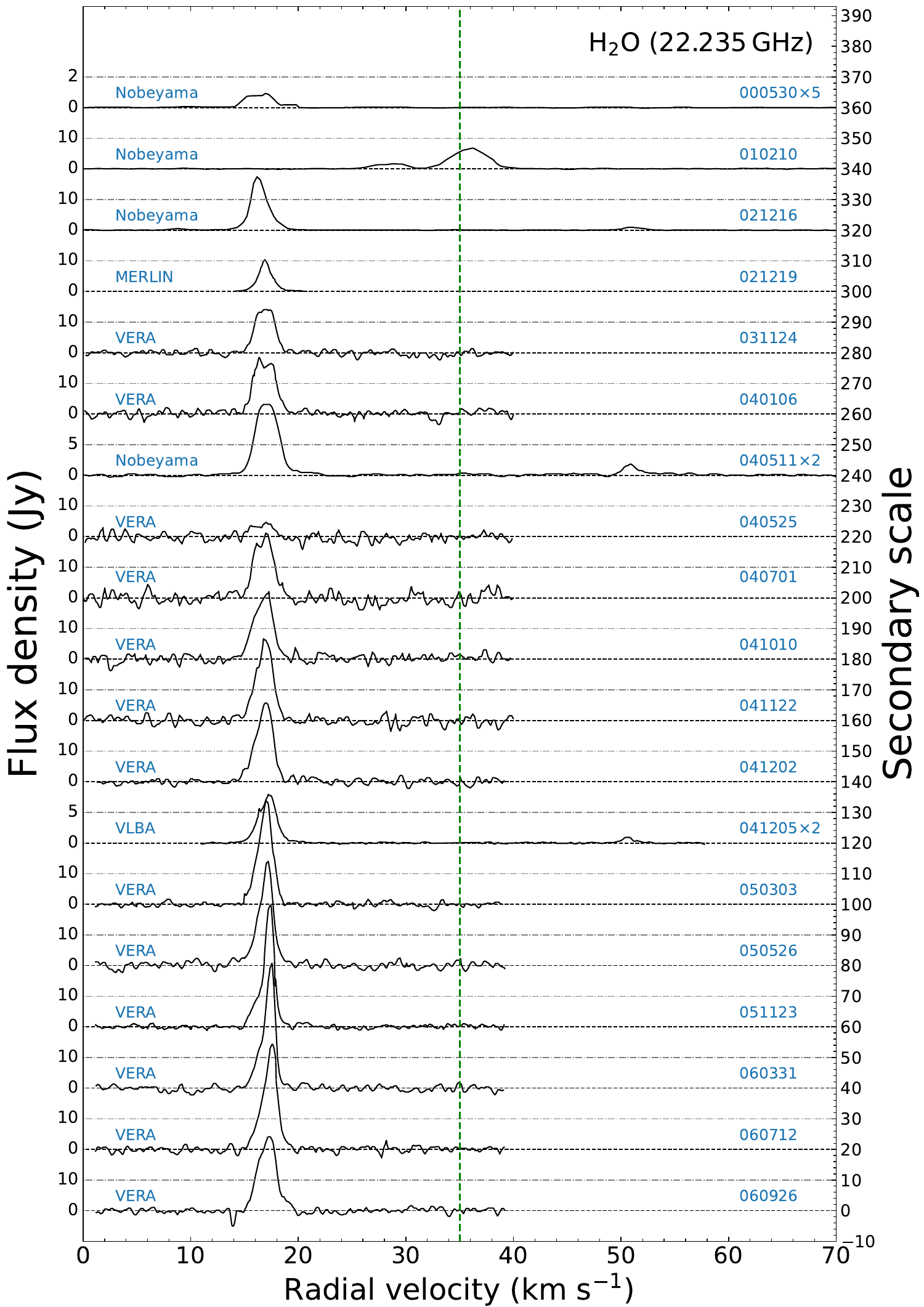}
    \caption{Time evolution of the H$_2$O 22.235\,GHz maser spectra of I19312 over the last 20 years. Nobeyama, MERLIN, VERA, KVN, and Effelsberg data were taken from \citet{2000PASJ...52L..43N, 2004PASJ...56.1083D, 2011ApJ...728...76N, 2008PASJ...60.1077S, 2007ApJ...669..446N, 2016JKAS...49..261K, 2013ApJ...769...20Y, 2014ApJ...794...81Y}, respectively. The date of observation (YYMMDD) is given for each spectral line. The intensity of the maser lines is given in Jy units, the conversion factors from Jy to K are 2.67\,Jy\,K$^{-1}$ \citep[]{2023PASJ...75.1183I}, 20\,Jy\,K$^{-1}$ \citep[]{2008PASJ...60.1077S} and 13.8\,Jy\,K$^{-1}$ \citep[]{2016JKAS...49..261K} for Nobeyama 45-m, VERA 20-m and KVN 21-m, respectively. The green dotted vertical line represents the systemic velocity of I19312 ($\sim$\,35\,km\,s$^{-1}$). To display the line flux densities, individual scales for each spectrum are provided on the left side in units of Jy. Additionally, a common secondary scale is shown on the right side to facilitate reading the absolute flux density values. }
    \label{fig: H2O-past-spectrum-1st}				
\end{figure}

The 22.235 GHz H$_2$O maser line has been observed seven times over the past 20 years excluding the current observation \citep{2000PASJ...52L..43N, 2007ApJ...669..446N, 2008PASJ...60.1077S, 2011ApJ...728...76N, 2013ApJ...769...20Y, 2014ApJ...794...81Y, 2016JKAS...49..261K}. Some of the past observations included monitoring observations over periods of 2 to 3 years \citep{2008PASJ...60.1077S, 2016JKAS...49..261K}. Figure~\ref{fig: H2O-past-spectrum-1st} and Figure~\ref{fig: H2O-past-spectrum-2nd}  show all the spectra obtained from past observations and the spectra obtained from the present observations in chronological order. From Figure~\ref{fig: H2O-past-spectrum-1st} and Figure~\ref{fig: H2O-past-spectrum-2nd}, it can be seen that, except for the detection of a line near the systemic velocity in February 2001, the line generally exhibits a double-peak profile. Additionally, from 2000 to 2012, a wider velocity interval double-peak profile (with relatively stronger blueshifted component) was observed, but it changed to a narrower velocity interval double-peak profile from around 2012 to 2015. The intensity of the 16--17 km~s$^{-1}$ component became very strong between 2004 and 2006, then gradually weakened, eventually disappearing completely on May 18, 2015. Interestingly, during the weakening process after 2006, the line intensity briefly increased again, showing a second, weaker peak around March 13, 2011.

\subsection{Observational Results of SiO Maser Lines and Comparison with Past Observations} \label{sec: sio}
Figures~\ref{fig: SiO42 and 43GHz} and \ref{fig: SiO86GHz} show the spectra of the SiO maser lines at 42.821 GHz, 43.122 GHz, and 86.243 GHz. Tables~\ref{tab: SiO42_43MHz parameters}, \ref{tab: SiO86.24GHz parameters}, and \ref{tab: SiO129.36GHz parameters} summarize the line parameters. The meanings of the parameters listed in the tables are the same as those in Table~\ref{tab: OH1612_65_67MHz parameters}.

The 42.821 GHz SiO maser line ($v=2, J=1-0$) was detected 10 out of 11 observation epochs, except for September 8, 2018. Basically, the line profile of this maser line is a single peak, with the radial velocity of the intensity peak confirmed to be around 55 km~s$^{-1}$. However, in the spectrum obtained on November 3, 2018, a weak feature appears to be present around 58--65 km~s$^{-1}$. During the observation period, the component at 55 km~s$^{-1}$ suddenly increased in flux density on May 16, 2018. Interestingly, the flux density of the 22.235 GHz H$_2$O maser line, as described in Section~\ref{sec: h2o}, also showed an increase in intensity around the same time (i.e., March 2018), though this was observed in the blueshifted component at 25 km~s$^{-1}$ (see Figure~\ref{fig: H2O22.2GHz}). The timescale of the intensity variation of this line is short, and the amplitude of the intensity variation is larger compared to other maser lines, as seen from the figure. After the sudden intensity increase on May 16, the intensity dropped below the detection limit by September 8, and then weak features reappeared about four months later. The 43.122 GHz SiO maser line ($v=1, J=1-0$) was detected in 8 out of the 11 observation epochs, excluding the first three epochs (see Figure~\ref{fig: SiO42 and 43GHz}). The radial velocity of the intensity peak was about 53 km~s$^{-1}$ (see Table~\ref{tab: SiO42_43MHz parameters}). The intensity of the 43.122 GHz SiO maser line also showed clear time variability during the observation period, particularly becoming stronger in the latter half of the observation period. Compared to the 42.821 GHz SiO maser line, the 43.122 GHz SiO maser line was relatively weak. It is noteworthy that when the intensity of the 42.821 GHz line suddenly increased on May 16, 2018, there was no significant change in the intensity of the 43.122 GHz line.

The 86.243 GHz SiO maser line ($v=1, J=2-1$) was detected in 5 observation epochs (see Figure~\ref{fig: SiO86GHz}). The radial velocity of the intensity peak varied from about 30 km~s$^{-1}$ to 52 km~s$^{-1}$. However, the detected features were very weak, and some of the emission peak detections were tentative, so only $F_{\rm peak}$, $V_{\rm peak}$, and $F_{\rm int}$ are listed in Table~\ref{tab: SiO86.24GHz parameters}. The 129.363 GHz SiO maser line ($v=1, J=3-2$) was not detected in this observation. Only the rms noise level is given in Table~\ref{tab: SiO129.36GHz parameters} for this line.

There have been 6 reported observations of SiO maser lines over the past 20 years \citep{2000PASJ...52L..43N,2004PASJ...56.1083D,2007ApJ...669..446N,2011ApJ...728...76N,2016JKAS...49..261K,2023A&A...669A.121Q}. Figures~\ref{fig: SiO42GHz-past-spectrum}, \ref{fig: SiO43GHz-past-spectrum}, and \ref{fig: SiO86GHz-past-spectrum} show the spectra obtained from past observations and the spectra obtained from this observation in chronological order. From the spectra observed around 2000 to 2001, it can be seen that emission components were detected on the blueshifted side relative to the systemic velocity for all lines. However, since 2002, it can be seen that emissions have been detected almost exclusively on the redshifted side (the 86 GHz line almost disappeared). The $J=1-0$ $v=1$ and $v=2$ lines often show a single peak on the redshifted side relative to the systemic velocity, but the radial velocity of the emission peak is not entirely constant and varies within a range of a few km~s$^{-1}$.

\subsection{Intercomparison of averaged spectra} \label{sec: comp}

Figure~\ref{fig: OH-H2O-SiO-comp} shows a comparison of the averaged spectra of the 1612 MHz OH maser line, the 42.821 GHz and 43.122 GHz SiO maser lines, and the 22.235 GHz H$_2$O maser line. For comparison, the spectrum of the CO $J=1-0$ line \citep{2005ApJ...633..282N} is also plotted in this figure. From past interferometric observations \citep{2005ApJ...633..282N} and model analyses \citep{2007A&A...470..957M,2023A&A...669A.121Q}, it is known that the molecular gas envelope of I19312 contains two kinematic components, known as the narrow component and the broad component (see Figure~\ref{fig:ShapeModel-CO}). The narrow component is a molecular flow that slowly moves in the bipolar direction, while the broad component is a small, approximately spherically symmetric expanding component. In the right panel of Figure~\ref{fig:ShapeModel-CO}, the CO line emission profiles of the narrow and broad components created from the model \citep{2023A&A...669A.121Q} are compared with the velocity ranges of the maser lines. From Figure~\ref{fig:ShapeModel-CO}, it is evident that the velocity range of the masers extends beyond the range that can be interpreted as the narrow component. Although the absolute positions of the maser line emission sources in I19312 have not been determined, from a velocity perspective, it can be inferred that the OH, H$_2$O, and SiO masers are likely associated with the broad component.

\section{Discussion} \label{sec:discussion}

When the SiO maser emission was first detected from I19312 in 2000, it was initially suggested that this object could be an AGB star or a post-AGB star \citep{2000PASJ...52L..43N}. However, subsequent observations revealed that I19312 could not be interpreted as a typical AGB star or post-AGB star. Instead, interpretations such as an evolved star with unique properties like OH~231.8+4.2 or W43A \citep{2011ApJ...728...76N}, an RSG \citep{2016ApJ...825...16N}, a YSO \citep{2016ApJ...828...51C}, or an RNR \citep{2011ApJ...728...76N} have been proposed. However, so far, no interpretation has been able to fully explain all the observational properties of I19312. In the following, based on previous discussions of the origin of I19312, we will use the results of our maser monitoring observations and previous maser observational data to see how far we can constrain the origin of I19312.  The discussion here will first summarize the issues of interpreting it as an evolved star with active mass loss (Section~\ref{sec: evolved stars}) and as a YSO (Section~\ref{sec: YSOs}). Then, we will  consider individually the objects that have some similarity to I19312 (Section~\ref{sec: other}). In reviewing the literature discussing the origin of I19312, it is clear that the fact that most known SiO maser sources are identified as evolved stars with active mass loss was the starting point for past discussions of the origin of I19312. Therefore, based on the previous discussions, we will first discuss the properties of SiO masers in each section, and then discuss the observational properties of other maser lines (and some additional observational properties).

\subsection{Comparison with Mass-Losing Evolved Stars} \label{sec: evolved stars}
At the time when the SiO maser emission line was detected from I19312 in 2000, most of the SiO maser sources found in the sky were considered to originate from evolved stars. This was the main reason why I19312 was initially suggested to be an AGB star or a post-AGB star. Indeed, by 2000, more than 1000 SiO maser sources were already known, but the only SiO maser sources clearly identified as objects other than evolved stars were the YSOs in three high-mass star-forming regions: Ori KL, W51 North, and Sgr B2 \citep{ 1975ApJ...197..329S, 1986mmmo.conf.....H}.

The SiO maser emission lines ($v=1\&2$ $J=1-0$; $v=1$ $J=2-1$) detected from I19312 with the Nobeyama 45~m telescope in 2000 showed relatively broad profiles centered around 20 km~s$^{-1}$ in the $v=1$ $J=1-0$ and $v=1$ $J=2-1$ lines. This kind of line profiles are occasionally seen in AGB stars. However, the $v=2$ $J=1-0$ line exhibited a double-peak profile with peaks around 25 km~s$^{-1}$ and 55 km~s$^{-1}$. Such double-peak profiles are relatively rare, except for the apparent double maser sources found in the densely packed star regions toward the Galactic center \citep{1999PASJ...51..355D}. A unique example, W43A, discussed in Section~\ref{sec: wf}, shows a similar double-peak profile \citep{2011ApJ...728...76N}. Figures~\ref{fig: SiO42GHz-past-spectrum}
 and~\ref{fig: SiO43GHz-past-spectrum} show all the spectra of the $v=1\&2$ $J=1-0$ lines observed in the past in chronological order. From these figures, it is clear that the double-peak line profile was observed only between around 2000 and 2004, and since then, the blue-shifted component has not been detected. Also, looking at the velocities of the detected emission peaks in the past, the SiO maser emission peaks of both the $v=1\&2$ $J=1-0$ lines are distributed over a velocity range of about 40 km~s$^{-1}$ from around 15 km~s$^{-1}$ to 55 km~s$^{-1}$. In the circumstellar envelopes of AGB stars, dust particles begin to condense at a distance of approximately 2 to 3 times the radius of the photosphere. The radiation from the central star exerts pressure on the dust particles, driving a spherical stellar wind. SiO masers are known to be emitted from regions near where the dust condenses \citep{2007IAUS..242..271B, 2009MNRAS.394...51G}. In these regions, the outward-moving dust particles accelerate the surrounding molecular gas, dragging it outward under radiation pressure. However, maser emission is not amplified in the direction of gas acceleration; instead, it is primarily beamed perpendicular to the acceleration. As a result, in spherically expanding circumstellar envelopes, the line-of-sight velocities of SiO maser intensity peaks are typically within 2 to 3 km~s$^{-1}$ of the systemic velocity of the central star \citep{1991A&A...242..211J}. The SiO maser emission lines of I19312 are sparsely detected over a wide velocity range of 40 km~s$^{-1}$. This observational fact is inconsistent with the properties of SiO masers emitted from spherically expanding envelopes (i.e., AGB star winds). SiO maser emission lines are also commonly detected from RSGs \citep{2012A&A...541A..36V}. In RSGs, the intensities of the $v=1$ $J=1-0$ and $v=1$ $J=2-1$ lines are generally brighter than the $v=2$ $J=1-0$ line \citep[see, e.g.,][]{2012A&A...541A..36V}. However, during the past 20 years, the SiO masers in I19312 have shown the larger intensity in $v=2$ $J=1-0$ than the $v=1$ $J=1-0$ and $v=1$ $J=2-1$ lines. Then, the SiO maser emission lines in RSGs often show line profiles with wider velocity widths compared to those of AGB stars due to the higher mass loss rate \citep{2012A&A...541A..36V}. And the SiO maser emission lines of RSGs are typically detected continuously over a wide velocity range \citep[see, e.g.,][]{2012ApJ...760...65F}. Again, the observational properties of the SiO masers of I19312 differ from those of RSGs. 

Next, we consider the properties of the OH maser lines. The 1612 MHz OH maser profile shown in Figure~\ref{fig: OH1612MHz} is clearly different from the line profile shown by spherically expanding envelopes. Unlike SiO masers, the 1612 MHz OH masers are distributed hundreds of times the radius of the photosphere \citep{1983ApJ...274..733B}. In such regions far from the central star, the acceleration mechanism due to radiation from the central star does not work, and the gas is considered to be moving outward at a constant terminal velocity \citep{1996A&ARv...7...97H}. In such situations, unlike SiO masers, maser amplification is possible in the radial direction. As a result, the 1612 MHz OH masers associated with spherically expanding envelopes often show double-peak line profiles with peaks at the red-shifted and blue-shifted sides of the envelope. However, it is also known that as the star evolves, the morphology and motion of the circumstellar envelope may deviate from spherically symmetric expansion, resulting in relatively irregular line profiles \citep{2004ApJS..155..595D}. Thus, the line profile of the OH maser line in I19312 is similar to that of an evolved star that is slightly more evolved than the AGB phase.

The intensity ratio of the three OH maser lines detected from I19312 tends to support the evolved star properties as mentioned in Section~\ref{sec:intro}. Comparing the emission intensities of the OH satellite line (i.e., the 1612 MHz line) and the main lines (the 1665 MHz and 1667 MHz lines), the intensity of the satellite line has exceeded that of the main lines throughout the observation period. OH maser sources are broadly classified into two types: Type I and II. In Type I, the main line intensity is stronger than the satellite line intensity, and in Type II, the satellite line intensity is stronger than the main line intensity. 

The OH main lines (at 1665 MHz and 1667 MHz) are produced by transitions between the hyperfine levels of the ground state of the OH molecule. It is widely accepted that these transitions are radiatively pumped, with the population inversion being driven by incoming radiation at wavelengths of 35 microns and 53 microns within the gas system \citep{1987ApJ...313..408D}. In addition to incoming continuum radiation, line overlap of molecules other than OH and collisional pumping could also be relevant. It is known that about 80\% of the evolved stars with both the main line and satellite line of OH detected simultaneously are Type II \citep[see, e.g.,][]{2000ApJ...533..959L}. Furthermore, to the best of our knowledge, there are currently no known OH maser sources other than evolved stars that exhibit solid Type II properties.  However, the pumping mechanisms of OH masers are not yet fully understood \citep[see, e.g.,][]{1996A&ARv...7...97H,2005NewA...10..283H} and it is difficult to completely rule out the possibility of Type II OH masers occurring in objects other than evolved stars, especially from objects with circumstellar environments similar to those of evolved stars. For example, it has recently been shown that Type II OH maser may be present in the YSOs \citep{2024PASA...41....9H}.


Regarding the H$_2$O maser, the wider double peak observed before 2012, which shifted to a narrower double peak after 2018, may be interpreted as the result of variations in the mass-loss process. Variations in the maser profile may trace changes in the conditions for maser excitation when density and/or velocity variations reach the maser emission regions.

\subsection{Comparison with Young Stellar Objects} \label{sec: YSOs}
One of the object groups where SiO maser emission lines are detected other than evolved stars is YSOs. As mentioned above, when SiO masers were detected from I19312 in 2000, SiO maser emission lines had been detected from three YSOs in high-mass star-forming regions. Over the following 20 years, the number of detected SiO maser emission from YSOs increased, and currently, SiO masers have been detected from a total of eight YSOs \citep{2016ApJ...826..157C}, all of which are YSOs in high-mass star-forming regions. SiO maser lines detected in evolved stars generally exhibit emission peaks within a very narrow velocity range near the systemic velocity \citep[typically within 2--3 km~s$^{-1}$,][]{1991A&A...242..211J}. In contrast, the spectrum of SiO maser lines detected in YSOs often shows emission peaks distributed over a wider velocity range than in the case of evolved stars, and it frequently exhibits irregular or double-peaked line profiles \citep{2009ApJ...691..332Z, 2016ApJ...826..157C}. In the case of SiO masers of I19312, as shown in Figures~\ref{fig: SiO42GHz-past-spectrum}
 and~\ref{fig: SiO43GHz-past-spectrum}, emission peaks appear at velocities away from the systemic velocity, and based on the SiO maser profile alone, it can be said that the characteristics are closer to those of YSOs rather than typical AGB stars (it should be noted that in the case of evolved stars, a double-peaked profile has also been detected in WF, as shown in Section~\ref{sec: wf}).

Regarding H$_2$O masers, in the case of YSOs, H$_2$O masers often accompany high-velocity molecular outflows, and it is sometimes observed that the detected velocity range exceeds 100 km~s$^{-1}$  \citep[see, Table 4 in][]{2024ApJS..270...13F}. However, even in evolved stars, the line width can exceed 100 km~s$^{-1}$ (see, Section~\ref{sec: wf}), so the line width of H$_2$O masers alone cannot be a decisive factor in determining whether an object is an evolved star or a YSO. As for OH masers, as mentioned in Section~\ref{sec: evolved stars}, to our knowledge, almost all OH masers detected from YSOs are currently Type I. In terms of OH maser properties, I19312 is closer to the properties of an evolved star than to those of a YSO.

All YSOs with detected SiO masers are associated with high-mass star-forming regions and have HII regions nearby \citep{2009ApJ...691..332Z, 2016ApJ...826..157C}. In the case of I19312, non-detection of near-infrared Br${\gamma}$ lines and centimeter-wave free-free emission has been confirmed \citep{2011ApJ...728...76N}. Additionally, as partly mentioned in Section~\ref{sec:intro} the infrared structure of I19312 is completely isolated and does not appear to be associated with a nearby star forming regions \citep{2000PASJ...52L..43N, 2007A&A...470..957M}. \citet{2007A&A...470..957M} estimated the stellar temperature of I19312 around 2800\,K, and concluded that the dust temperature is too low for that of a high-mass YSO \citep{1999ApJ...525..772P}. Additionally, \citet{2004PASJ...56.1083D} and \citet{2023A&A...669A.121Q} have pointed out that single-dish radio spectroscopic observations suggest a high abundance ratio of $^{13}$C in I19312. Due to the issue of optical thickness, there is uncertainty in measuring the carbon isotope ratio from carbon-bearing radio emission lines, but if the dominance of $^{13}$C is true, it would be a strong factor against the YSO hypothesis. Furthermore, the distance to I19312 also places strong limitations on the YSO hypothesis. Assuming the distance obtained from the annual parallax of the H$_2$O maser source by VLBI \citep{2011PASJ...63...81I}, the bolometric flux corresponds to about $\rm 21500_{-6900}^{+9700}$ solar luminosities (see, Appendix~\ref{sec:sed}). This luminosity is generally considered too bright for the absolute luminosity of a low-mass YSO \citep{1994AAS...185.4815H, 2004ApJ...617.1177W}.  In conclusion, although more SiO maser sources have recently been found in YSOs, we do not believe that YSO can explain the nature of I19312.

\subsection{Comparison with other types of objects}\label{sec: other}
Regarding the origin of I19312, early discussions primarily focused on whether it was an evolved star or a YSO. However, as the observational properties of I19312 have gradually become clearer, specific objects and/or groups of specifitc objects that share more common characteristics with I19312, aside from typical evolved stars and YSOs, have come to be considered as possible interpretations. Below, we will organize the current situation on this issues, including new possibilities beyond the previously discussed specific objects and object groups, and discuss considerations related to the origin of I19312.

\subsubsection{OH 231.8+4.2} \label{sec: OH 231.8}
OH 231.8+4.2 (Rotten Egg Nebula) is known as a proto-planetary nebula \citep{2002A&A...389..271B}. This object has been noted as potentially the same type as I19312 due to the properties shared with I19312, such as the simultaneous detection of radio molecular lines probing both carbon-rich and oxygen-rich chemical environments \citep{2000A&A...357..651S}, the detection of SiO maser lines \citep{2019MNRAS.488.1427K}, and the association with extended infrared nebulosity \citep{2001A&A...373..932A, 2019MNRAS.488.1427K}. However, the central star of this object is a binary system \citep{2019MNRAS.488.1427K}, and the SiO maser is believed to be emitted from the circumstellar envelope of a mira variable (QX Pup) within the binary system. The thermal line profiles of the SiO molecules ($v=0$ $J=2-1$)  detected from OH 231.8+4.2 show parabolic profiles typical of AGB stars \citep{2019MNRAS.488.1427K}. For I19312, the results of the maser monitoring observations, as shown in the figures in Section~\ref {sec: results}, do not display the time variations typically associated with mira-type variable stars. Furthermore, the WISE photometric data, collected over a period of more than 3000 days, shows a constant luminosity within the limits of photometric accuracy, further supporting the conclusion that I19312 is not a mira-type variable star. In this regard, the properties of the SiO maser sources in I19312 and OH 231.8+4.2 fundamentally differ. However, the SiO maser from OH 231.8+4.2 shows a line profile with 2--3 brightness peaks within a 20 km~s$^{-1}$ range centered around the systemic velocity, which differs from the typical line profile of SiO masers emitted from spherically expanding shells \citep{2019MNRAS.488.1427K}. This unique line profile suggests that the circumstellar envelope of the mira star within the binary system deviates from a spherically symmetric expansion pattern, possibly influenced by the gravitational field (Roche lobe) of the binary system. If the central star of I19312 forms a binary system, a similar mechanism could potentially affect the SiO maser profile of I19312. 

The 1667 MHz OH maser line of OH 231.8+4.2 is brighter than the 1612 MHz OH maser line, exhibiting the characteristics of a Type I OH maser source. These OH lines are detected over a wide velocity range, extending up to 100 km~s$^{-1}$ \citep{1980AJ.....85..724M}. \citet{1980AJ.....85..724M} proposed that a non-local infrared radiation field at 80 $\mu$m is responsible for exciting the 1667 MHz emission. This may suggest a difference in the excitation conditions of the OH maser gas between I19312 and OH 231.8+4.2. The 1667 MHz line profile shares similarities with I19312, such as the absence of the double-peaked profile typically seen in evolved stars and its detection over a broad velocity range. \citet{1982ApJ...259..625M} reported, based on observations with VLA, that the emission source of the 1667 MHz OH maser line forms a ring-like structure, suggesting that the emission originates from the interaction region between the AGB wind and the high-velocity outflow.

\subsubsection{Water Fountains} \label{sec: wf}
Water Fountains are evolved stars in the transition from the end of the AGB phase to the post-AGB phase, during which the shape of the circumstellar envelope changes from spherical to non-spherical \citep{2023PASJ...75.1183I}. It is suggested that SiO masers around post-AGB stars rapidly disappear \citep{1989ApJ...338..234L}. However, WF are believed to be stars that have just entered or are about to enter the post-AGB phase, and indeed, weak SiO masers have been detected from W43A \citep[see, e.g.][]{2005ApJ...622L.125I}, a proto-typical WF \citep[Note: The disappearance of the SiO maser in W43A has been reported.][]{2022AJ....163...85A}. The SiO maser profile of W43A is double-peaked, similar to the line profile of I19312's SiO maser \citep[see, e.g.,][]{2005ApJ...622L.125I}. Recently, SiO masers have also been detected from another WF, IRAS 16552$-$3050 \citep{2022AJ....163...85A}. The SiO maser profile of IRAS 16552$-$3050 is single-peaked, but the peak velocity is about 15 km~s$^{-1}$ redshifted compared to the systemic velocity obtained from CO radio observations. The fact that the peak velocity of the SiO maser deviates from the systemic velocity is common among W43A, IRAS 16552-3050, and I19312. Not all known WFs show SiO maser lines, but this is expected if we consider that WFs are objects in the transition phase from the AGB phase to the post-AGB phase.

The characteristic of WF is that there is a high-velocity molecular bipolar outflow in the center of the circumstellar envelope, mainly traced by H$_2$O masers. Since the star has just entered the post-AGB phase, a spherically symmetric AGB wind still remains in the outer envelope, and the outer envelope is traced by the OH 1612 MHz maser line. Therefore, the velocity range where 1612 MHz OH masers are detected can serve as a probe for the expansion velocity of the AGB wind. Thus, comparing the velocity ranges of the 1612 MHz OH masers and the 22.235 GHz H$_2$O masers, the high-velocity jets associated with H$_2$O masers are often observed outside the velocity range of the OH masers. This phenomenon is used to search for WF candidates \citep{2024ApJS..270...13F}. In the case of I19312, the H$_2$O maser emission are detected outside the velocity range of the OH maser emission, making it a candidate for WF \citep{2011ApJ...728...76N}. Indeed, based solely on the maser properties, the possibility that I19312 is a WF cannot be ruled out.

However, the WF hypothesis for I19312 faces issues in explaining, for example, the large molecular gas mass inferred from CO observations \citep{2016ApJ...825...16N} and the complex chemical composition \citep{2004PASJ...56.1083D, 2023A&A...669A.121Q} as mentioned in Section~\ref{sec:intro}.  From an evolutionary perspective, the SiO masers in WF are expected to disappear over time. According to the present results,  the SiO maser in I19312 has not disappeared for at least the last 20 years. Monitoring the presence of the SiO maser in I19312 in the future is meaningful for constraining the WF hypothesis.

\subsubsection{U Equ} \label{sec: u equ} 
U Equ is a peculiar evolved star that has changed its spectral type from M-type to F-type over about 30 years \citep[see, e.g.,][]{2024A&A...682A.133K}. In visible spectroscopic observations conducted in the 1990s \citep{1996A&A...310..259B}, the spectrum of U Equ showed many molecular absorption lines, typical of M-type spectra. However, in the latest observations \citep{2024A&A...682A.133K}, the molecular absorption lines have mostly disappeared, and neutral and ionized molecular emission lines dominate. Due to the very rapid spectral change, this object is thought to be transitioning from the late AGB phase to the post-AGB phase. Maser lines of SiO (43.122 GHz), H$_2$O (22.235 GHz), and OH (1612 MHz, 1667 MHz) have been detected from U Equ \citep{2024A&A...682A.133K}. The 43.122 GHz SiO maser shows a single peak profile near the systemic velocity, similar to the typical SiO maser profile of AGB stars. The 1612 MHz OH maser shows a double-peak profile with a velocity separation of about 5--6 km~s$^{-1}$, and the 22.235 GHz H$_2$O maser is detected outside the velocity range of the 1612 MHz OH maser. Therefore, according to the criteria of \citet{2024ApJS..270...13F}, it is classified as a WF candidate. The complexity of the maser profiles also shares similarities with I19312. However, no extended infrared nebula, as seen with I19312, has been detected from U Equ.

\subsubsection{HD 101584} \label{sec: hd101584} 
In early discussions of I19312, one reason it was difficult to interpret I19312 as an evolved star was the detection of methanol lines \citep{2004PASJ...56.1083D, 2015PASJ...67...95N}. However, since \citet{2017A&A...603L...2O} detected methanol from HD 101584, a post-AGB star, the interpretation of a methanol-accompanied evolved star cannot be ruled out now. HD 101584 is considered to be in the late AGB phase or early post-AGB phase. After completing the AGB phase, the increase in UV radiation from the central star and the interaction of fast stellar winds with slower stellar winds often give post-AGB star circumstellar envelopes unique chemical compositions \citep{2020ChJCP..33..668M}. Indeed, several molecular spieces, including CO, SiO, SO, CS, H$_2$CO, and CH$_3$OH, have been detected from this objects, which is a similarity with I19312. The issue is whether a circumstellar maser like I19312 can exist in a star like HD 101584. Currently, only the 1667 MHz OH maser has been detected from HD 101584 \citep{1992ApJ...390L..23T}, and no other maser lines have been reported \citep{2019A&A...623A.153O}. The detection of the 1667 MHz OH maser without the detection of the 1612 MHz OH maser is unusual for an evolved star. Thus, focusing on the nature of the OH maser, it can be said that the properties of I19312, a Type II OH maser source, and HD 101584, only with its 1667 MHz OH maser line, are somewhat different.

\subsubsection{V407 Cyg} \label{sec: v407 cyg} 
Another unique circumstellar SiO maser source worth mentioning is V407 Cyg. V407 Cyg is particularly known as a symbiotic star \citep{2013AASP....3...75D}. This object attracted attention due to a nova explosion on March 10, 2010. V407 Cyg is a binary system, with one star being a white dwarf and the other a mira variable (i.e., AGB star). SiO masers have been detected from V407 Cyg both before and after the nova explosion \citep{2011PASJ...63..309D}. Monitoring observations after the 2010 nova explosion confirmed dramatic changes in the SiO maser spectrum \citep{2015JKAS...48..267C}. Specifically, high-velocity components disappeared a few weeks after the explosion, followed by the appearance of low-velocity components. This change is interpreted as the shock wave of the nova passing through the maser emission region, temporarily disrupting the maser conditions while changing the gas velocity, and later, the gas in the same region cooling down and re-emitting the SiO maser. Additionally, the emission intensity ratio of the $v=1$ and $v=2$ $J=1-0$ lines changed, with the $v=1$ $J=1-0$ line being stronger before the explosion and the $v=2$ $J=1-0$ line becoming relatively stronger after the explosion. 

Currently, there is no evidence supporting that I19312 is a binary system or a symbiotic star. However, if I19312 has experienced some explosive event, it is conceivable that the physical state and velocity of the SiO maser emission region could change, altering the line profile and emission intensity, similar to V407 Cyg's SiO maser. In fact, the time variation of the line profile shown by the SiO maser of I19312 is reminiscent of the time variation of the SiO maser of V407 Cyg.

\subsubsection{Red nova remnants} \label{sec: rnr} 
Objects where both methanol and SiO masers lines are detected are very important comparative targets for interpreting I19312. In the case of HD 101584 mentioned in Section~\ref{sec: hd101584}, while it is similar to I19312 in that methanol lines are detected, SiO masers are not detected. On the other hand, the RNR may satisfy both conditions. Among the five confirmed RNRs so far, SiO masers have been detected from V838 Mon \citep{2005PASJ...57L..25D}. V838 Mon experienced a nova explosion (merger event) in January 2002. V838 Mon is also known for its light echo images observed by the Hubble Space Telescope \citep[HST,][]{2008AJ....135..605S}. The SiO maser of V838 Mon shows a stronger intensity in the $v=2$ $J=1-0$ line than in the $v=1$ $J=1-0$ line, which is similar to I19312. However, the line profile is a simple single-peak profile with a peak velocity near the systemic velocity, differing from I19312. Although there are no examples of SiO maser and methanol lines being detected simultaneously from the same RNR, methanol  lines have been detected from another RNR, CK Vul \citep{2017A&A...607A..78K}. It is known that the nature of RNRs varies greatly depending on the properties of the progenitor stars before the merger \citep{2024arXiv240103919K}. The conditions under which both methanol and SiO maser detection coexist are currently unknown. However, the fact that there is one RNR each where both methanol and SiO maser lines are detected provides important clues for considering RNRs as a possible origin of I19312.

\section{Summary} \label{sec:summary}

In this study, we conducted long-term monitoring observations of SiO, H$_2$O, and OH maser lines for the IRAS source 19312+1950 (I19312) using single-dish radio telescopes. Additionally, we compiled all observational results of SiO, H$_2$O, and OH maser lines for I19312 conducted over the past approximately 20 years and compared the characteristics of I19312's masers with those of similar objects to consider the origin of I19312. The main findings and conclusions are as follows:

\begin{enumerate} 
\item In all three OH maser transitions, the maser features exhibited non-periodic variations over the 8-month monitoring period, with amplitude changes by a factor of up to about 2. The average flux density was approximately 5 Jy km~s$^{-1}$ at 1612 MHz, about 1 Jy km~s$^{-1}$ at 1665 MHz, and about 1.5 Jy km~s$^{-1}$ at 1667 MHz. The velocity ranges for all three transitions were similar to those observed in 2000, 2003, and 2012, except for the extension of the 1612-MHz velocity range due to the detection of a new 38 km~s$^{-1}$ feature. The OH maser emission of I19312 shows an intensity ratio greater than 1 between the satellite line at 1612 MHz and the main lines (1665/1667 MHz), a characteristic commonly observed in evolved stars. The line profiles suggest that if the source is a mass-losing evolved star, it is in the post-AGB phase.

\item During approximately 1.5 years of monitoring the 22.235 GHz H$_2$O maser line, we observed a change in the profile. Stronger emission was detected at 22--28 km~s$^{-1}$ until January 2019, after which it shifted to 38--42 km~s$^{-1}$. The maser variations were more pronounced than those seen for OH, with a brightness increase by a factor of more than 30 for the 22--28 km~s$^{-1}$ feature. The total velocity range (22--42 km~s$^{-1}$) was significantly narrower during 2018-–2020 compared to the range observed between 2000 and 2015, where strong emission (up to 40.8 Jy) was seen near 17 km~s$^{-1}$, and weaker emission was detected at velocities greater than 50 km~s$^{-1}$. In comparison, much larger brightness variations were observed during 2004–-2006. The H$_2$O maser emission was detected over a wider velocity range than the OH maser, suggesting the presence of molecular outflows. This characteristic is also observed in YSOs, but it also makes I19312 a potential candidate for a WF in post-AGB stars.

\item SiO maser emission was detected between 2018 and 2020 at 42.8 and 43.1 GHz, with velocities ranging from 50 to 55 km~s$^{-1}$. The brightness of the maser features fluctuated by a factor of a few. Tentative detections of maser features at levels below 0.1 Jy were made at 86.2 GHz, while no features were observed at 129.4 GHz above 0.1 Jy. The velocity range over which SiO maser emission has been detected in the past 20 years has changed significantly. Overall, the SiO $v = 2$, $J=1-0$ (42.821 GHz) emission has now been observed across a broad range of 20~km~s$^{-1}$$< v < 55$~km~s$^{-1}$. While such a wide SiO maser line profile resembles those detected from some massive YSOs, it is unusual for evolved stars, where emission typically occurs near the systemic velocity and spans less than 10 km~s$^{-1}$. Although the SiO maser line profile alone shows similarities with those from massive YSOs, we remain skeptical about classifying I19312 as a YSO due to many other inconsistencies.

\item The velocities and evolution of the maser features for all three maser species suggest a complex geometry for the emission regions. In relation to the two kinematic components predicted by a model of the CO emission, the velocity ranges indicate that the maser emission regions are likely associated with the broader component, which represents a spherical outflow in the inner part of I19312's circumstellar environment.

\item We derived a stellar luminosity of $21500_{-6900}^{+9700}$~L$_{\odot}$ using the bolometric flux from the spectral energy distribution between 0.5 and 870~$\mu$m and the distance determined from a VLBI annual parallax measurement. If I19312 is an evolved star, this luminosity is typical of an AGB star with an intermediate mass of $>2$ M$_{\odot}$.

\item After comparing all observational properties with those of YSOs,
evolved stars, Red Nova remnants, and several peculiar stars, we
conclude that the origin and evolutionary stage of I19312 remain
unclear.

\end{enumerate}

To place stronger constraints on the origin of I19312 from the perspective of its maser properties, future research will need to (1) conduct regular observations of the SiO, H$_2$O, and OH masers to identify the emergence of bright phases; (2) use these phases for interferometric observations to determine the relative positions of the SiO, H$_2$O, and OH maser emissions; and (3) investigate the molecular gas components and their relationship with the observational characteristics of maser lines.

\begin{acknowledgments}
We acknowledge the science research grants from the China Manned Space Project with No. CMS-CSST-2021-A03, No.CMS-CSST-2021-B01. 
JN acknowledges financial support from the `One hundred top talent program of Sun Yat-sen University' grant no. 71000-18841229.  The KVN observations were made possible by the high-speed network connections among the KVN sites provided by the Korea Research Environment Open NETwork (KREONET), which is managed and operated by the Korea Institute of Science and Technology Information (KISTI). The work described in this paper was partially supported by the Major Project Research Fund (2018–2020) of the Korea Astronomy and Space Science Institute (KASI). YZ thanks the financial supports from the National Natural Science Foundation of China (NSFC, No.12473027 and No.12333005) and the Guangdong Basic and Applied Basic Research Funding. JQ thanks the financial supports from the National Natural Science Foundation of China (NSFC, No. 12463006).
\end{acknowledgments}

\appendix

\section{Updated infrared spectral energy distribution for I19312} \label{sec:sed}

 The infrared total flux of I19312 has been calculated in previous studies \citep{2004PASJ...56..193N, 2011ApJ...728...76N, 2016ApJ...825...16N}. However, with the recent release of numerous infrared archive data, the number of available infrared photometric data has significantly increased compared to the time when the stellar luminosity was previously calculated. Therefore, the stellar luminosity of I19312 is recalculated here using the most up-to-date dataset. The photometry data were collected using the VizieR Catalogue\footnote{http://vizier.cds.unistra.fr/vizier/sed/}. The collected infrared photometry data are summarized in Table~\ref{tab: SED of I19321}. Additionally, the spectral energy distribution plot is provided in Figure~\ref{fig: NIR_SED}. The calculation of the luminosity is based on the distance ($\rm 3.8_{-0.58}^{+0.83}$ kpc) derived from the annual parallax measured with the VLBI technique \citep{2011PASJ...63...81I}. The annual parallax measurement is based on the proper motion of the 22.235 GHz H$_2$O maser source in I19312. By integrating the spectral energy distribution over the wavelength range from 0.55 to 870~$\mu$m using a trapezoidal approximation, an absolute luminosity of approximately $\rm 21500_{-6900}^{+9700}~L_{\odot}$ was obtained.

\input{Target-and-Observed-Lines}
\input{obs-date}
\input{OH1612_65_67_BasicParameter}
\input{OH1720_BasicParameter}
\input{H2O_blue_red_BasicParameter.tex}
\input{SiO42_43_BasicParameter.tex}
\input{SiO86p24_BasicParameter.tex}
\input{SiO129p36_BasicParameter.tex}

\addtocounter{figure}{-1} 

\begin{figure}
  \centering
  \renewcommand{\thefigure}{9b}
    \includegraphics[width=0.7\textwidth, angle=0]{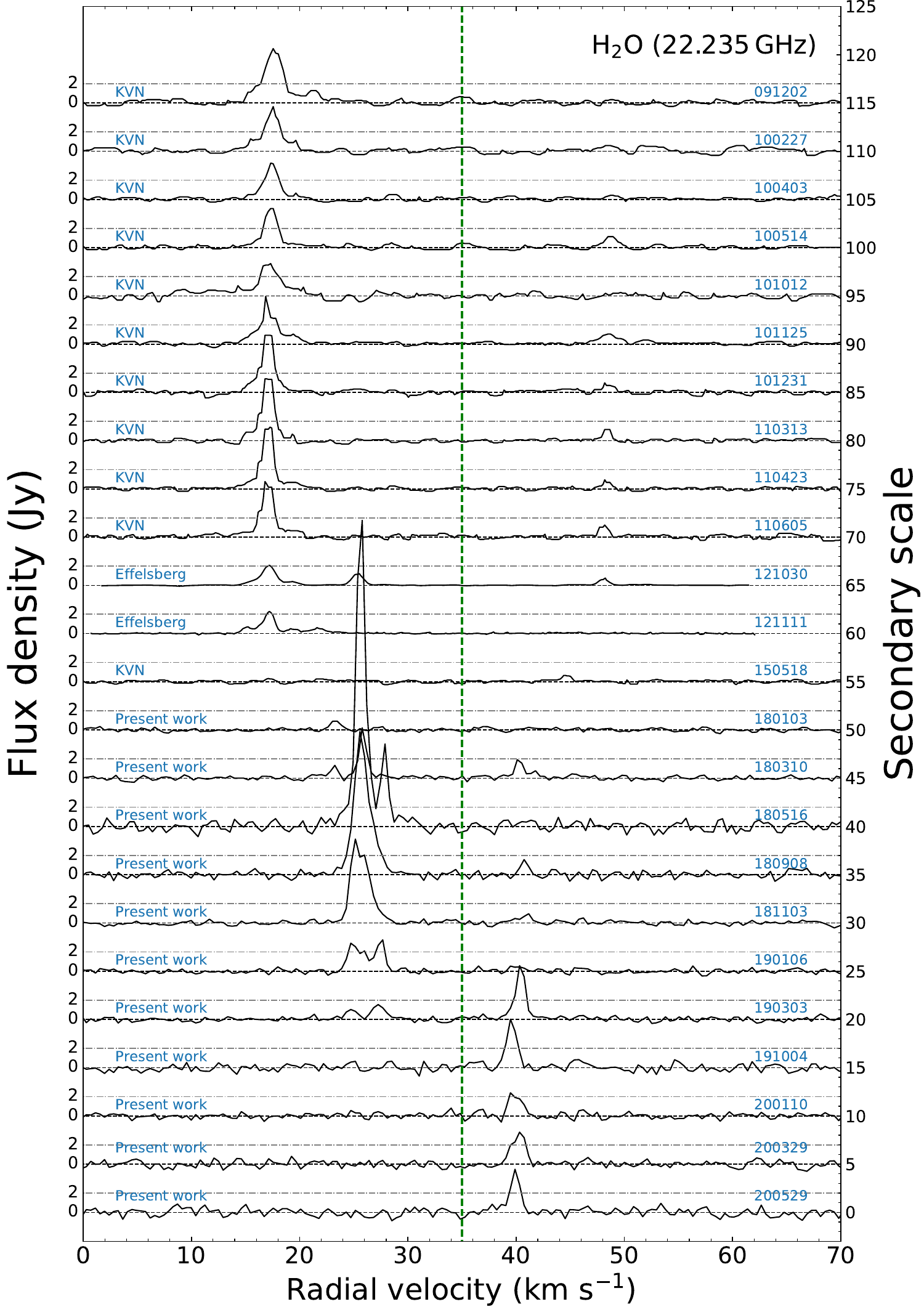}
\caption{Time evolution of the H$_2$O 22.235\,GHz maser spectra of I19312 over the last 20 years (continued).}
\label{fig: H2O-past-spectrum-2nd}
\end{figure}

\begin{figure}
    \centering
    \includegraphics[width=0.6\textwidth]{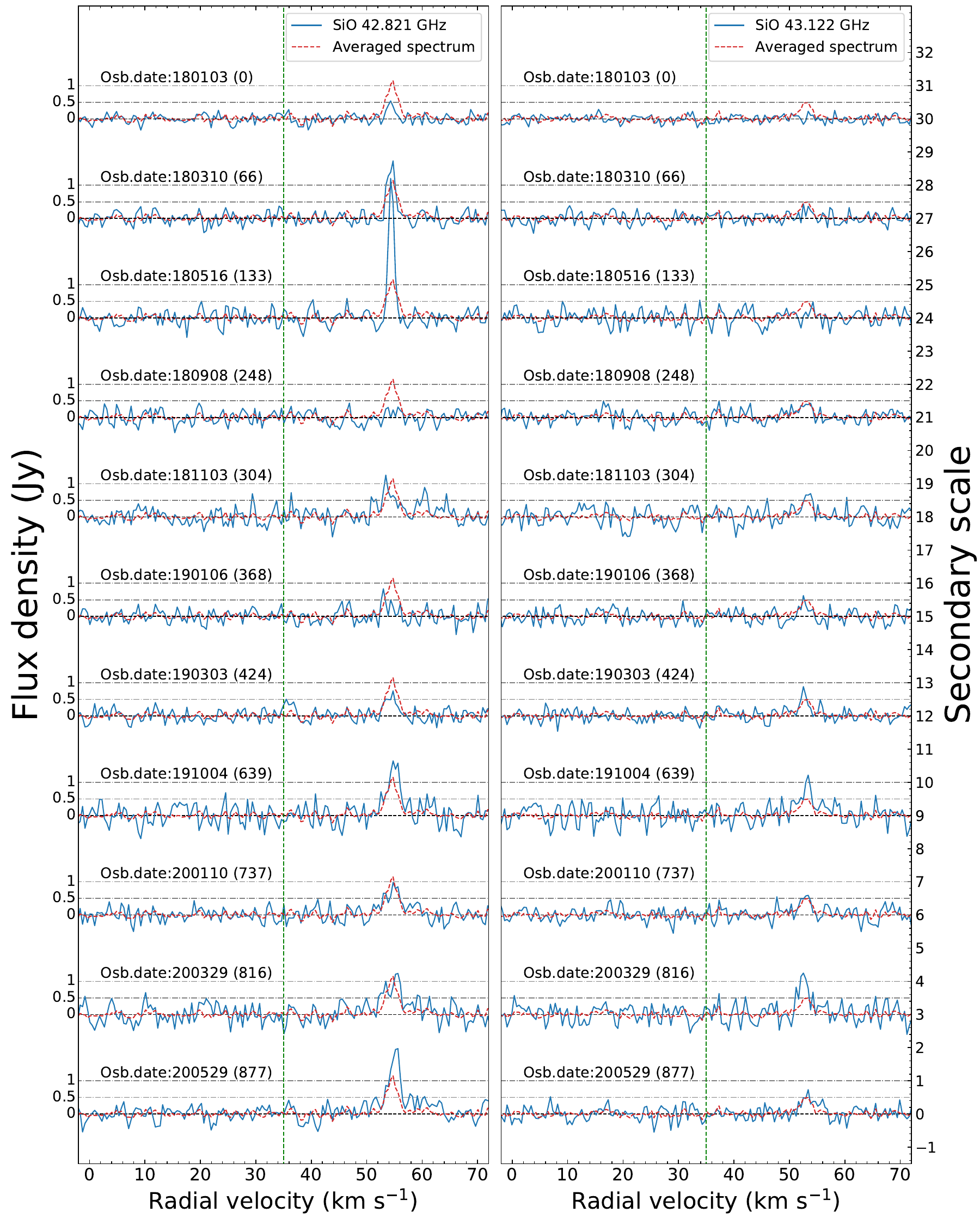}
    \caption{Spectra of the SiO\,42.821\,GHz and SiO\,43.122\,GHz maser lines of I19312 from 2018 January to 2020 May. Notations are the same as in Figure~\ref{fig: OH1612MHz}.}
    \label{fig: SiO42 and 43GHz}				
\end{figure}

\begin{figure}
    \centering
    \includegraphics[width=0.6\textwidth]{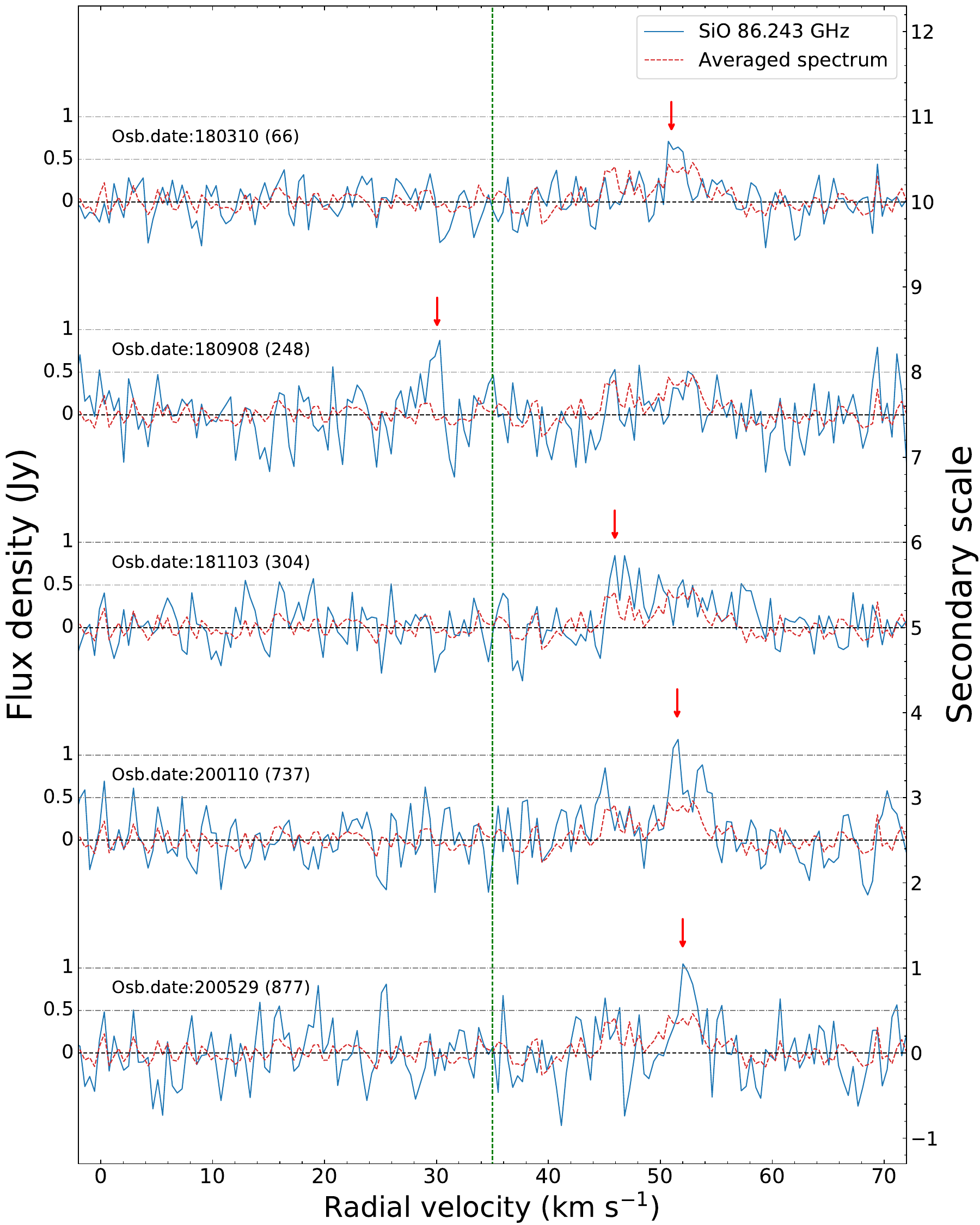}
    \caption{Selected spectra of the SiO\,86.243\,GHz line of I19312 from 2018 January to 2020 May. The red arrows represent emission peaks that were considered likely to be detected. Other notations are the same as in Figure~\ref{fig: OH1612MHz}.}
    \label{fig: SiO86GHz}				
\end{figure}

\begin{figure}
    \centering
    \includegraphics[width=0.7\textwidth]{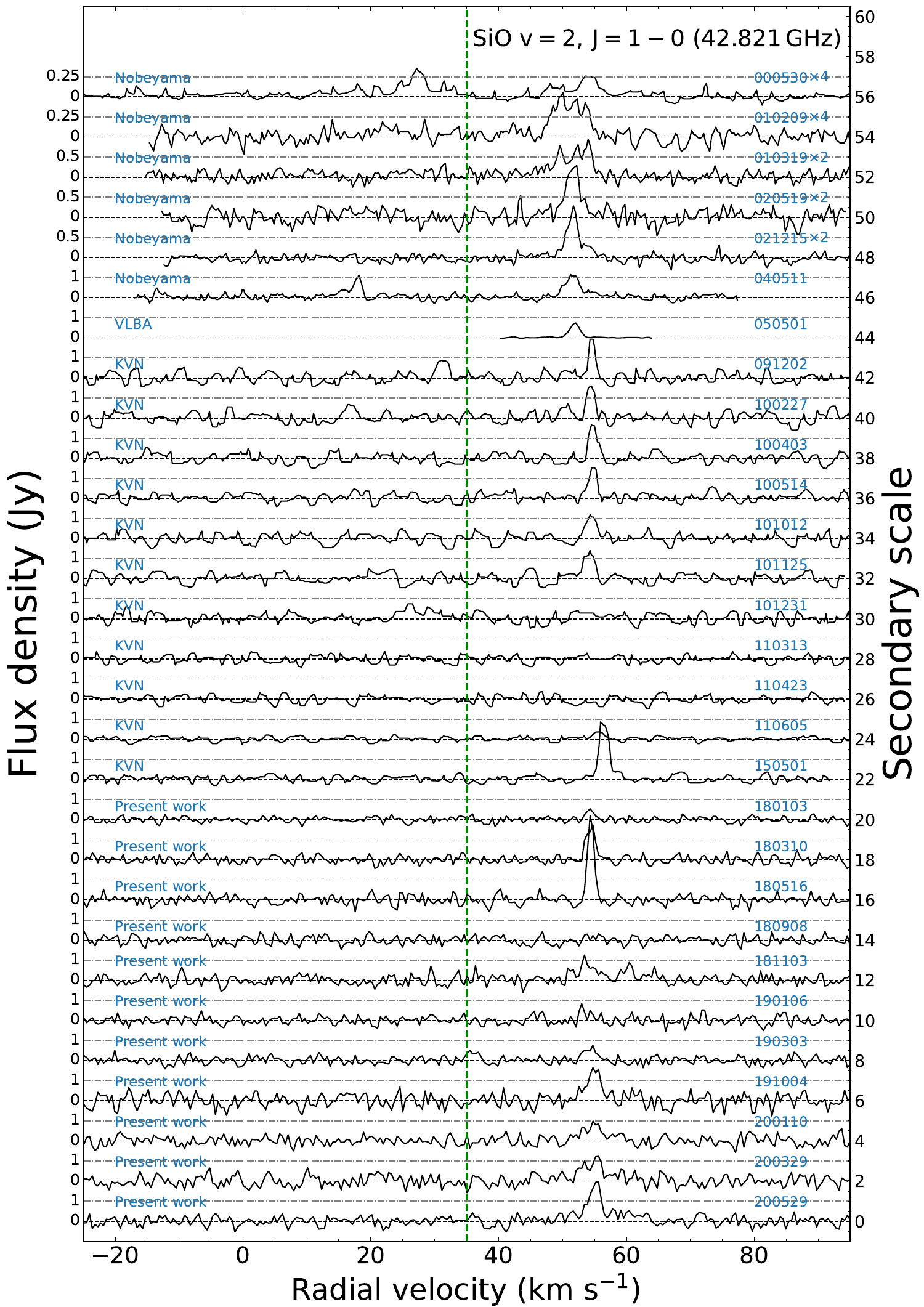}
    \caption{Time evolution of the SiO\,42.821\,GHz maser spectra of I19312 over the last 20 years. The data were collected from the literature (see text). Nobeyama, VLBA and KVN data were taken from \citet{2000PASJ...52L..43N, 2004PASJ...56.1083D, 2007ApJ...669..446N, 2011ApJ...728...76N, 2016JKAS...49..261K}. The data of observation (YYMMDD) is given for each spectral line. The conversion factors from Jy to K are 3.2\,Jy\,K$^{-1}$ \citep[]{2023PASJ...75.1183I} and 13.1\,Jy\,K$^{-1}$ \citep[]{2016JKAS...49..261K} for Nobeyama 45-m and KVN 21-m, respectively. The green dotted vertical line represents the systemic velocity of I19312 ($\sim$\,35\,km\,s$^{-1}$). Past spectra are magnified along the vertical axis to view line profiles. Magnification is given to the right of the observation date. To display the line flux densities, individual scales for each spectrum are provided on the left side in units of Jy. Additionally, a common secondary scale is shown on the right side to facilitate reading the absolute flux density values. }
    \label{fig: SiO42GHz-past-spectrum}				
\end{figure}

\begin{figure}
    \centering
    \includegraphics[width=0.7\textwidth]{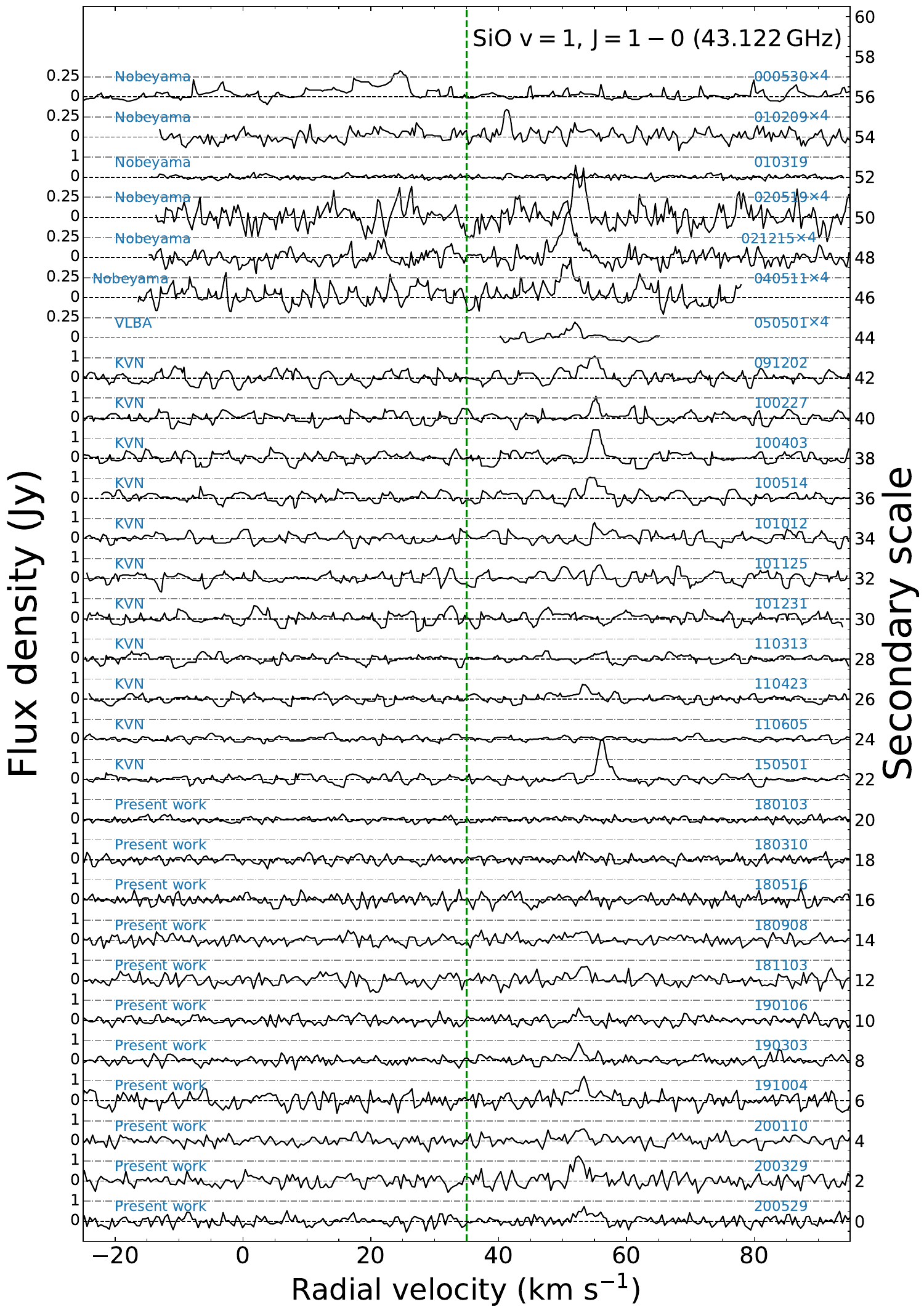}
    \caption{Time evolution of the SiO\,43.122\,GHz maser spectra of I19312 over the last 20 years. Nobeyama, VLBA and KVN data were taken from \citet{2000PASJ...52L..43N, 2004PASJ...56.1083D, 2007ApJ...669..446N, 2011ApJ...728...76N, 2016JKAS...49..261K}. The data of observation (YYMMDD) is given for each spectral line. The conversion factors from Jy to K are 3.2\,Jy\,K$^{-1}$ \citep[]{2023PASJ...75.1183I} and 13.1\,Jy\,K$^{-1}$ \citep[]{2016JKAS...49..261K} for Nobeyama 45-m and KVN 21-m, respectively. The green dotted vertical line represents the systemic velocity ($\sim$\,35\,km\,s$^{-1}$). Past spectra are magnified along the vertical axis to view line profiles. Magnification is given to the right of the observation date. To display the line flux densities, individual scales for each spectrum are provided on the left side in units of Jy. Additionally, a common secondary scale is shown on the right side to facilitate reading the absolute flux density values. }
    \label{fig: SiO43GHz-past-spectrum}				
\end{figure}

\begin{figure}
    \centering
    \includegraphics[width=0.7\textwidth]{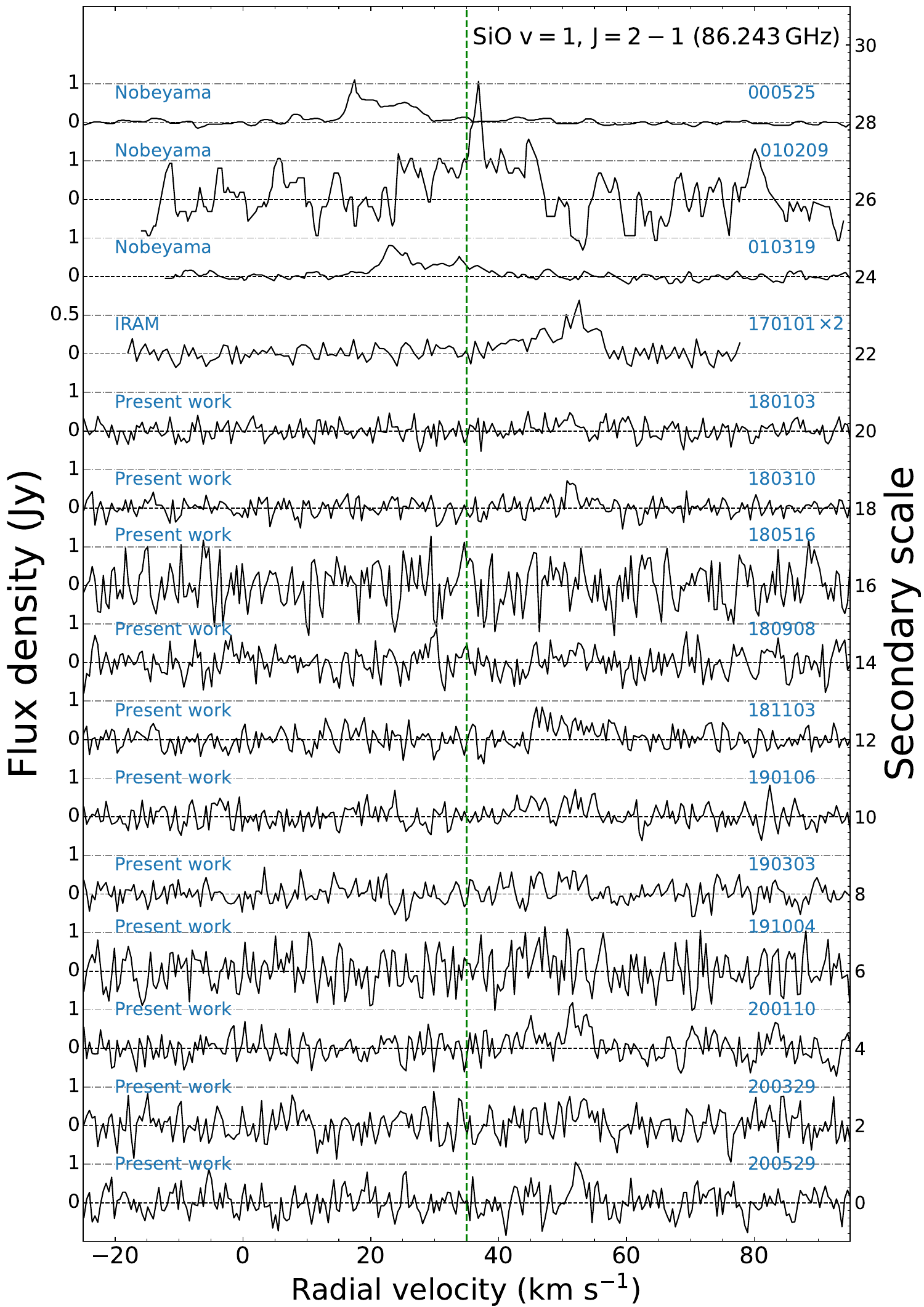}
    \caption{Time evolution of the SiO\,86.243\,GHz maser spectra of I19312 over the last 20 years. Nobeyama and IRAM data were taken from  \citet{2000PASJ...52L..43N, 2004PASJ...56.1083D, 2023A&A...669A.121Q}.  The data of observation (YYMMDD) is given for each spectral line. The conversion factors from Jy to K are 4.0\,Jy\,K$^{-1}$ \citep[]{2007ApJ...669..446N}, 6.2\,Jy\,K$^{-1}$ \citep[]{2003ASSL..283..363M} and 15.9\,Jy\,K$^{-1}$ \citep[]{2016JKAS...49..261K} for Nobeyama 45-m, IRAM 30-m and KVN 21-m, respectively. The green dotted vertical line represents the systemic velocity ($\sim$\,35\,km\,s$^{-1}$). Past spectra are magnified along the vertical axis to view line profiles. Magnification is given to the right of the observation date. To display the line flux densities, individual scales for each spectrum are provided on the left side in units of Jy. Additionally, a common secondary scale is shown on the right side to facilitate reading the absolute flux density values. }
    \label{fig: SiO86GHz-past-spectrum}				
\end{figure}

\begin{figure}
    \centering
    \includegraphics[width=0.6\textwidth]{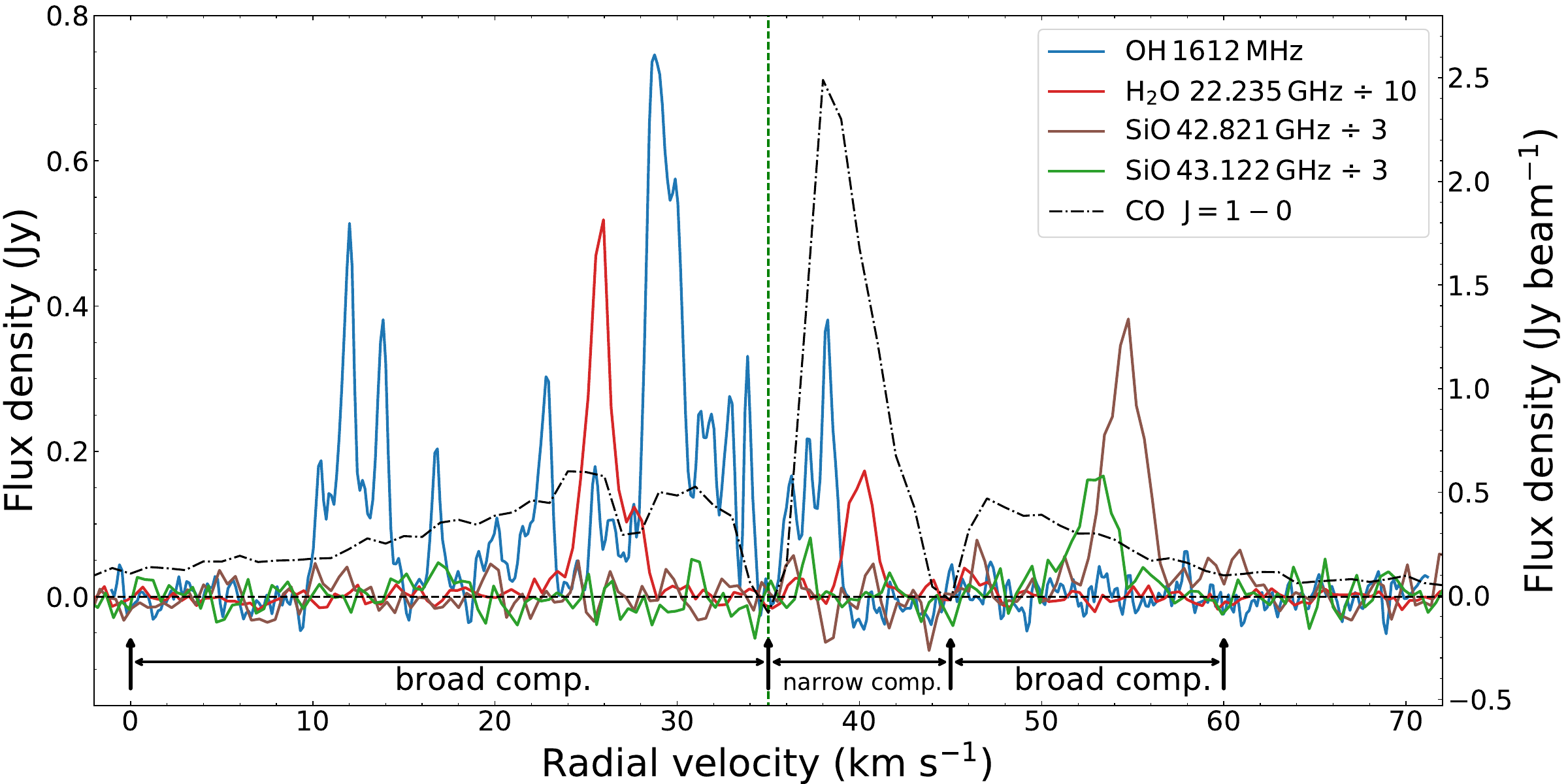}
    \caption{Comparison of the averaged OH\,1612\,MHz, H$_2$O\,22.235\,GHz, SiO\,43.122\,GHz and SiO\,42.821 GHz maser spectra. The CO $J=1-0$ spectrum obtained with BIMA \citep{2005ApJ...633..282N} is superimposed for comparison. The intensity of the maser lines is given in Jy units (see left vertical axis) and that of the CO line is given in Jy\,beam$^{-1}$ units (see right vertical axis). The velocity range of the narrow and broad components is given at the bottom (see, text).}
    \label{fig: OH-H2O-SiO-comp}			
\end{figure}

\begin{figure}
  \centering
  \subfigure{\includegraphics[width=0.4\textwidth]{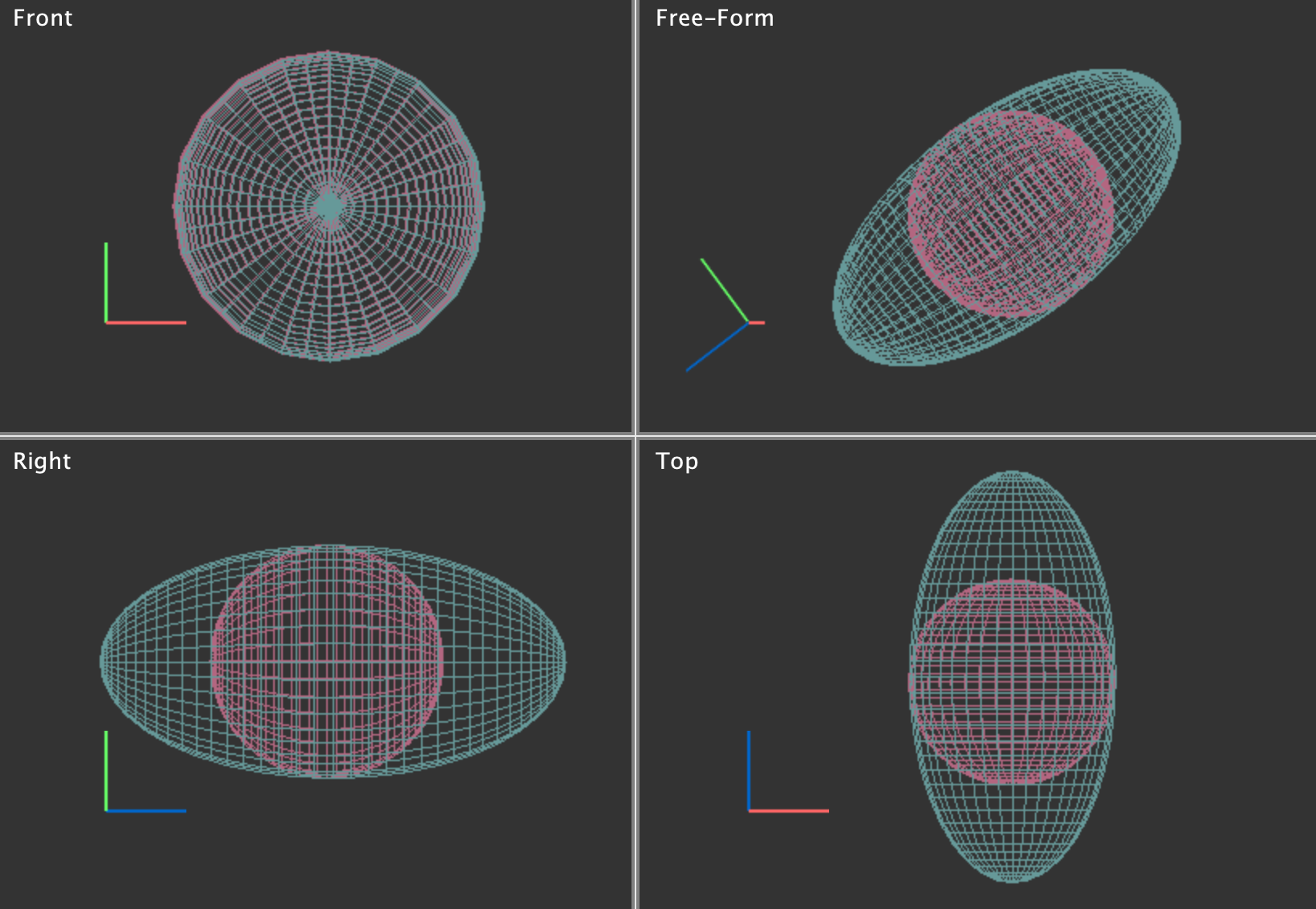}}
  \hspace{2em}
  \subfigure{\includegraphics[width=0.45\textwidth]{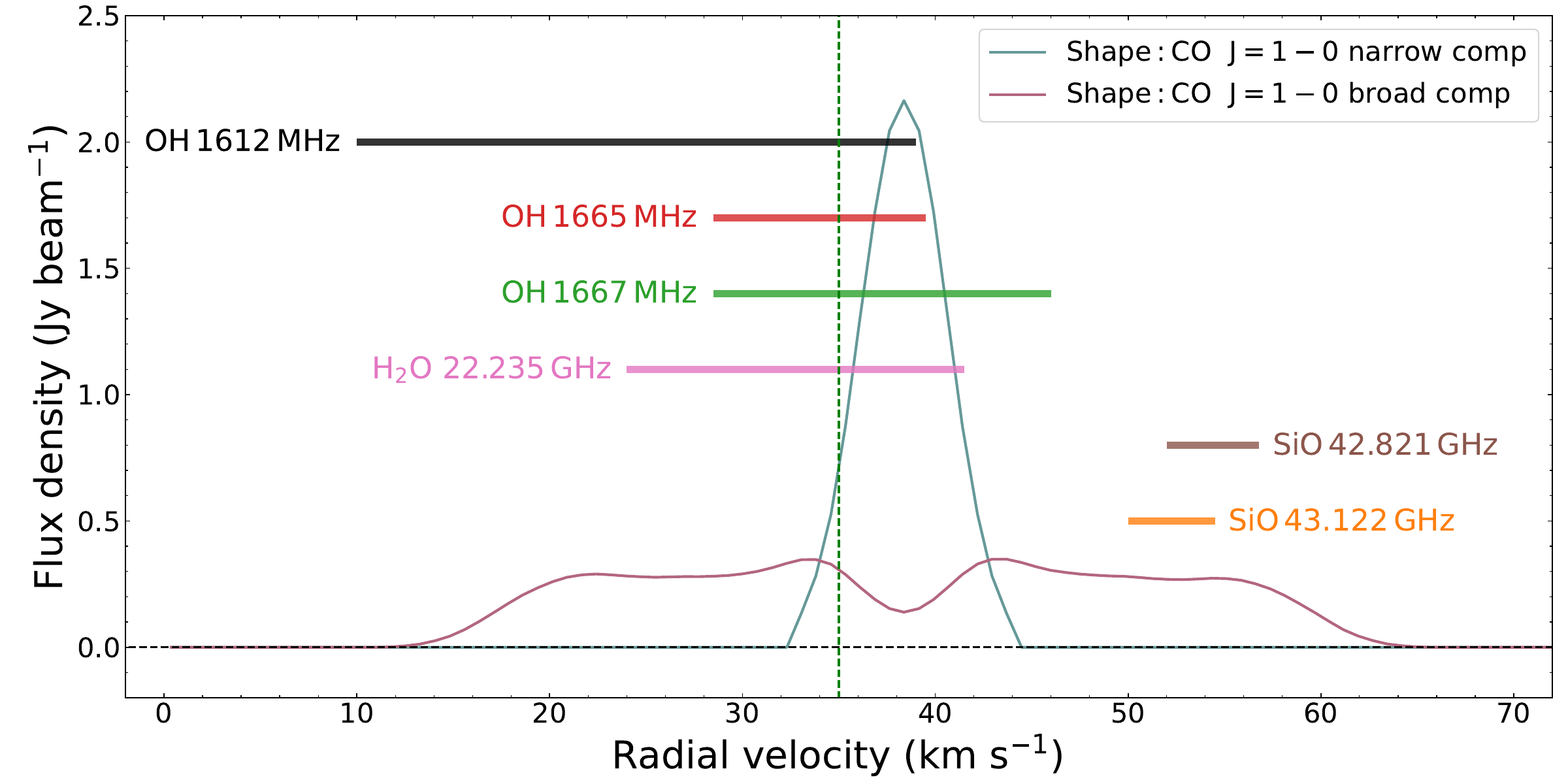}}
  \caption{Left panel: Model geometry of the Shape model from four different angles \citep{2023A&A...669A.121Q}. The model was built based on the distribution of CO gas. The Free-Form view means the observer's view with a position angle of $-37\degree$, where up is north and left is east. The modeled structures are composed of an inner spherical envelope (light red), an outer elliptical envelope (light blue). Right panel: a comparison between the Shape simulated CO $J=1-0$ spectra and the velocity ranges of the averaged spectra for the three maser species lines during the present observation period.}
  \label{fig:ShapeModel-CO}
\end{figure}

\begin{figure}
	\centering
	\includegraphics[width=0.6\textwidth]{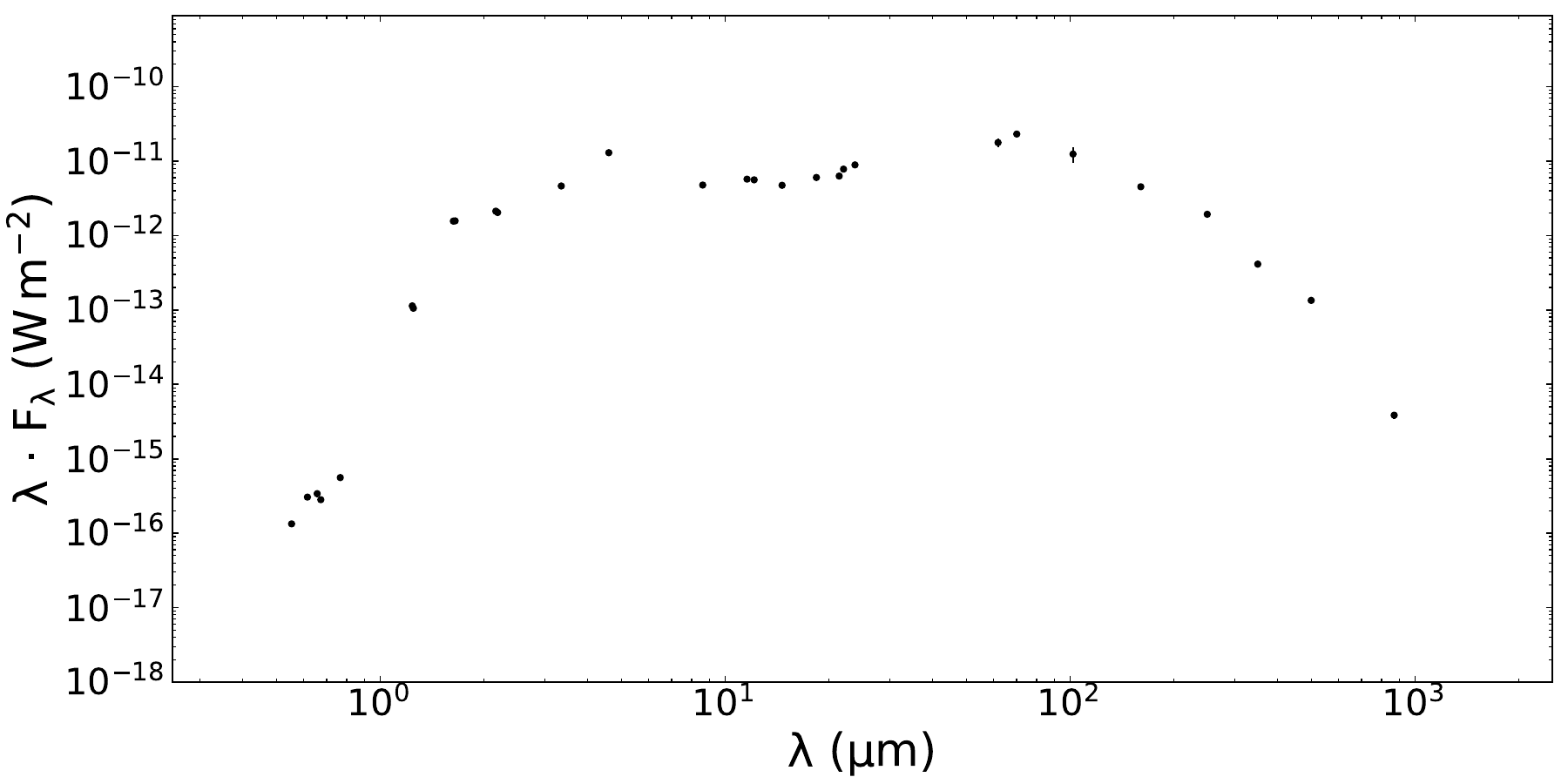}
	\caption{Spectral energy distribution of IRAS 19312+1950 in the optical and infrared wavelengths based on the latest data set. See Appendix~\ref{sec:sed} for details.}
	\label{fig: NIR_SED}					
\end{figure}

\input{SED_Data_Param_SingleColum.tex}

\bibliography{ms}{}
\bibliographystyle{aasjournal}

\end{CJK*}
\end{document}

%% file: Target-and-Observed-Lines.tex
\begin{deluxetable*}{ccccc}
	\tablecaption{Target Position and Observed Lines.}
	\label{tab:Target and Observed Lines}
	\tablewidth{0pt}
	\tablehead{\colhead{Target source} & \colhead{R.A.(J2000)}  & \colhead{Dec.(J2000)}   & Note \\ 
		& (hh mm ss.ss)                            &(dd mm ss.s)                            &   } 
	\startdata       
	IRAS 19312+1950  & 19 33 24.25                   & +19 56 55.7    &  2MASS position  \\    
	\\       
	\hline
	\hline
	Molecule  &  Transition &  Frequency\,(GHz) &  Velocity Coverage\,(\,km\,$\rm s^{-1}$)  &Telescope \\
	\hline
	\multirow{4}{*}{OH}  &  $^2$\uppercase\expandafter{\romannumeral2}$_{3/2}\,J=3/2\,F=1-2$  & 1.612   & $-1.0$ $\sim$ 71.2  &NRT  \\
	&  $^2$\uppercase\expandafter{\romannumeral2}$_{3/2}\,J=3/2\,F=1-1$   & 1.665    & 0.1  $\sim$ 70.0   &NRT  \\
	&  $^2$\uppercase\expandafter{\romannumeral2}$_{3/2}\,J=3/2\,F=2-2$  & 1.667    & 0.2  $\sim$ 70.0  &NRT \\
	&  $^2$\uppercase\expandafter{\romannumeral2}$_{3/2}\,J=3/2\,F=2-1$  & 1.721    &  1.2  $\sim$ 68.9  &NRT  \\
	\hline               
	H$_2$O  & $6_{1,6} - 5_{2,3}$   & 22.235     &  $-179.2$  $\sim$ 251.4 &KVN  \\    
	\hline      
	\multirow{4}{*}{SiO}  & $v=1, J=1-0$ &  43.122 & $-29.6$ $\sim$ 192.0 &KVN  \\    
	& $v=2, J=1-0$ & 42.821     &  $-124.6$  $\sim$ 98.5  &KVN  \\ 
	& $v=1, J=2-1$ & 86.243     &  $-74.9$  $\sim$ 146.8  &KVN  \\   
	& $v=1, J=3-2$ & 129.363     & $-37.8$  $\sim$ 109.9  &KVN \\
	\enddata
\end{deluxetable*}

%% file: obs-date.tex
\begin{deluxetable}{ c c c c c c c c} 
\caption{Date of observations.}
\label{tab: obs date}
\tablehead{	\multicolumn{3}{c}{KVN\,(SiO, H$_2$O)}  & & & \multicolumn{3}{c}{NRT\,(OH)}\\
\cline{1-3} \cline{6-8}
Date   & Julian Date & TD & &  & Date & Julian Date & TD \\
(yyyy-mm-dd)   &  &  & &  & (yyyy-mm-dd)   &  &  }
\startdata
2018-01-03&2458122&0       & & & 2018-02-02&2458152&30\\
2018-03-10&2458188&66    & & & 2018-03-30&2458208&86\\
2018-05-16&2458255&133  & & & 2018-05-02&2458241&119\\
2018-09-08&2458370&248 & & & 2018-07-28&2458328&206\\
2018-11-03&2458426&304  & & & 2018-07-29&2458329&207\\
2019-01-06&2458490&368 & & & 2018-09-17&2458379&257\\
2019-03-03&2458546&424 & & & 2018-09-18&2458380&258\\
2019-10-04&2458761&639  & & & 2018-09-30&2458392&270\\
2020-01-10&2458859&737  & & & ... & ... & ... \\
2020-03-29&2458938&816 & & & ... & ... & ... \\
2020-05-29&2458999&877 & & & ... & ... & ... \\
\enddata
\tablecomments{TD: Total number of days since the start of monitoring observations (January 3, 2018).}
\end{deluxetable}

%% file: OH1612_65_67_BasicParameter.tex
\begin{deluxetable}{rrrrrrrrrr} 
	\caption{Summary of OH line observations.}
	\label{tab: OH1612_65_67MHz parameters}
	\tablehead{Line & Date	&	$V_{\rm min}$	&	$V_{\rm max}$	&	$V_{\rm range}$	&	$F_{\rm peak}$	&	$V_{\rm peak}$	&	$F_{\rm int}$	&	$\rm rms$	&	$\rm SNR$	\\
		&	(yyyy-mm-dd)	&	($ \rm \,km\ s^{-1}$)	&	($ \rm \,km\ s^{-1}$)	&	($ \rm \,km\ s^{-1}$)	&	(Jy)	&	($ \rm \,km\ s^{-1}$)	&	($ \rm Jy \,km\ s^{-1}$ )	&	(Jy) & }
	
	\startdata
	\multirow{8}{*}{\makecell{OH\\1612\,MHz}}	&	2018-02-02	&	10.60 	&	38.83 	&	28.23 	&	0.92 	&	28.62 	&	6.76 	&	0.05 	&	20.05 	\\
	&	2018-03-30	&	11.59 	&	38.40 	&	26.81 	&	0.47 	&	28.76 	&	2.76 	&	0.04 	&	10.89 	\\
	&	2018-05-02	&	10.17 	&	38.97 	&	28.80 	&	0.97 	&	27.62 	&	7.05 	&	0.03 	&	28.41 	\\
	&	2018-07-28	&	10.32 	&	38.69 	&	28.37 	&	0.81 	&	28.76 	&	4.08 	&	0.04 	&	21.57 	\\
	&	2018-07-29	&	10.32 	&	38.69 	&	28.37 	&	0.68 	&	28.76 	&	3.34 	&	0.04 	&	16.25 	\\
	&	2018-09-17	&	11.45 	&	38.97 	&	27.52 	&	0.80 	&	28.90 	&	3.86 	&	0.05 	&	17.72 	\\
	&	2018-09-18	&	11.59 	&	38.40 	&	26.81 	&	0.81 	&	28.62 	&	4.29 	&	0.06 	&	13.65 	\\
	&	2018-09-30	&	11.31 	&	30.18 	&	18.87 	&	0.85 	&	29.04 	&	4.01 	&	0.10 	&	8.21 	\\
	\hline
	\multirow{8}{*}{\makecell{OH\\1665\,MHz}}   &	2018-02-02	&	29.51 	&	38.71 	&	9.20 	&	0.28 &	32.25 	&	1.80 	&	0.04 	&	6.97 \\
	&	2018-03-30	&	…	&	…	&	…	&	$^{*}$0.14 	&	$^{*}$58.07 	&	$^{*}$0.14 	&	0.04 	&	3.37 	\\
	&	2018-05-02	&	29.37 	&	38.16 	&	8.79 	&	0.23 	&	32.53 	&	1.40 	&	0.03 	&	6.95 	\\
	&	2018-07-28	&	29.23 	&	38.98 	&	9.75 	&	0.22 	&	29.92 	&	1.31 	&	0.04 	&	5.96 	\\
	&	2018-07-29	&	… 	&	… 	&	… 	&	$^{*}$0.24 	&	$^{*}$30.06 	&	$^{*}$1.46 	&	0.05 	&	4.77 	\\
	&	2018-09-17	&	29.51 	&	39.26 	&	9.75 	&	0.25 	&	29.78 	&	1.53 	&	0.04 	&	6.87 	\\
	&	2018-09-18	&	… 	&	… 	&	… 	&	$^{*}$0.29 	&	$^{*}$32.67 	&	$^{*}$1.50 	&	0.07 	&	4.36 	\\
	&	2018-09-30	&	…	&	…	&	…	&	$^{*}$0.41 	&	$^{*}$36.79 	&	$^{*}$1.20 	&	0.13 	&	3.17 	\\
	\hline
	\multirow{8}{*}{\makecell{OH\\1667\,MHz}}	&	2018-02-02	&	29.24 	&	43.92 	&	14.68 	&	0.24 	&	34.73 	&	1.18 	&	0.04 &	5.77 	\\
	&	2018-03-30	&	…	&	…	&	…	&	$^{*}$0.16 	&	$^{*}$35.96 	&	$^{*}$0.42 	&	0.04 	&	3.61 	\\
	&	2018-05-02	&	33.08 	&	45.43 	&	12.35 	&	0.23 	&	34.59 	&	1.49 	&	0.03 	&	6.95 	\\
	&	2018-07-28	&	28.69 	&	36.65 	&	7.96 	&	0.28 	&	34.45 	&	1.35 	&	0.03 	&	8.38 	\\
	&	2018-07-29	&	33.22 	&	45.01 	&	11.80 	&	0.29 	&	34.86 	&	2.02 	&	0.05 	&	6.03 	\\
	&	2018-09-17	&	30.47 	&	45.15 	&	14.68 	&	0.29 	&	34.45 	&	1.52 	&	0.04 	&	8.25 	\\
	&	2018-09-18	&	… 	&	… 	&	… 	&	$^{*}$0.36 	&	$^{*}$35.27 	&	$^{*}$1.97 	&	0.10 	&	3.51 	\\
	&	2018-09-30	&	… 	&	… 	&	… 	&	$^{*}$0.40 	&	$^{*}$33.90 	&	$^{*}$2.16 	&	0.08 	&	4.81 	\\
	\enddata
 \tablecomments{As a result of visually inspecting the data, we identified features that fall below 5$\sigma$ but are likely to be detected, and these are marked with an asterisk (*).}
\end{deluxetable}

%% file: OH1720_BasicParameter.tex
\begin{deluxetable}{cll} 
	\caption{Summary of OH 1720 MHz line observations.}
	\label{tab: OH1720_MHz parameters}
	
	\tablehead{Line	&	Date	&	$\rm rms$	\\						
		&	(yyyy-mm-dd)	& (Jy) }	
	\startdata
	\multirow{8}{*}{\makecell{OH\\1720\,MHz}}& 2018-02-02	& 0.04	\\
	&2018-03-30	&	0.04	\\
	&2018-05-02	&	0.04	\\
	&2018-07-28 &   0.04    \\
	&2018-07-29	&	0.05	\\
	&2018-09-17 & 0.03	\\
	&2018-09-18  &	0.16	\\
	&2018-09-30&	0.17	\\
	\enddata
\end{deluxetable}

%% file: H2O_blue_red_BasicParameter.tex
\begin{deluxetable}{crrrrrrrrr} 
	\caption{Summary of H$_2$O maser line observations.}
	\label{tab: H2O_blue_red parameters}	
	
	\tablehead{Line & Date & $V_{\rm min}$	& $V_{\rm max}$	& $V_{\rm range}$ &	$F_{\rm peak}$ & $V_{\rm peak}$	& $F_{\rm int}$	& $\rm rms$	& $\rm SNR$ \\
		&	(yyyy-mm-dd)	&	($ \rm \,km\ s^{-1}$)	&	($ \rm \,km\ s^{-1}$)	&	($ \rm \,km\ s^{-1}$)	&	(Jy)	&	($ \rm \,km\ s^{-1}$)	&	($ \rm Jy \,km\ s^{-1}$ )	&	(Jy) &  }	
	
	\startdata
	\multirow{11}{*}{\makecell{H$_2$O\\22.235\,GHz}}&	2018-01-03	&	23.04 	&	23.89 	&	0.84 	&	0.97 	&	23.47 	&	2.76 	&	0.14 	&	6.93 	\\
	&	2018-03-10	&	22.83 	&	40.53 	&	17.70 	&	5.24 	&	25.78 	&	10.90 	&	0.14 	&	37.43 	\\
	&	2018-05-16	&	24.10 	&	28.31 	&	4.21 	&	31.74 	&	25.78 	&	47.61 	&	0.41 	&	77.41 	\\
	&	2018-09-08	&	24.31 	&	27.68 	&	3.37 	&	14.90 	&	25.57 	&	28.29 	&	0.28 	&	53.21 	\\
	&	2018-11-03	&	24.31 	&	27.68 	&	3.37 	&	8.69 	&	25.15 	&	20.29 	&	0.14 	&	62.07 	\\
	&	2019-01-06	&	24.31 	&	28.10 	&	3.79 	&	3.31 	&	27.68 	&	8.97 	&	0.14 	&	23.64 	\\
	&	2019-03-03	&	24.31 	&	41.16 	&	16.85 	&	5.52 	&	40.32 	&	12.70 	&	0.28 	&	19.71 	\\
	&	2019-10-04	&	39.05 	&	40.32 	&	1.26 	&	4.97 	&	39.48 	&	8.97 	&	0.28 	&	17.75 	\\
	&	2020-01-10	&	39.48 	&	40.74 	&	1.26 	&	2.35 	&	39.48 	&	4.83 	&	0.28 	&	8.39 	\\
	&	2020-03-29	&	39.48 	&	40.74 	&	1.26 	&	3.31 	&	40.32 	&	4.97 	&	0.28 	&	11.82 	\\
	&	2020-05-29	&	39.48 	&	40.32 	&	0.84 	&	4.42	&	39.90 	&	5.24	&	0.41	&	10.78	\\
	\hline
	\multirow{11}{*}{\makecell{H$_2$O\\22.235\,GHz\\(Blue-shifted component)}}& 	2018-01-03	&	23.04 	&	23.89 	&	0.84 	&	0.97 	&	23.47 	&	1.38 	&	0.14 	&	6.93  	\\
	&	2018-03-10	&	22.83 	&	26.63 	&	3.79 	&	5.24 	&	25.78 	&	7.04 	&	0.14 	&	37.43 	\\
	&	2018-05-16	&	24.10 	&	28.31 	&	4.21 	&	31.74 	&	25.78 	&	46.09 	&	0.41 	&	77.41 	\\
	&	2018-09-08	&	24.31 	&	27.68 	&	3.37 	&	14.90 	&	25.57 	&	26.91 	&	0.28 	&	53.21 	\\
	&	2018-11-03	&	24.31 	&	27.68 	&	3.37 	&	8.69 	&	25.15 	&	18.08 	&	0.14 	&	62.07 	\\
	&	2019-01-06	&	24.31 	&	28.10 	&	3.79 	&	3.31 	&	27.68 	&	8.28 	&	0.14 	&	23.64 	\\
	&	2019-03-03	&	24.31 	&	28.10 	&	3.79 	&	1.52 	&	27.26 	&	4.28	&	0.28 	&	5.43 	\\
	&	2019-10-04	&	…	&	…	&	…	&	…	&	…	&	…	&	0.28 	&	…	\\
	&	2020-01-10	&	…	&	…	&	…	&	…	&	…	&	…	&	0.28 	&	…	\\
	&	2020-03-29	&	…	&	…	&	…	&	…	&	…	&	…	&	0.28 	&	…	\\
	&	2020-05-29	&	…	&	…	&	…	&	…	&	…	&	…	&	0.41	&	…	\\
	\hline
	\multirow{11}{*}{\makecell{H$_2$O\\22.235\,GHz\\(Red-shifted component)}}&	 2018-01-03 	&	…	&	…	&	…	&	…	&	…	&	…	&	0.14 	&	…	\\
	&	2018-03-10	&	40.11 	&	40.53 	&	0.42 	&	1.93 	&	40.11 	&	0.27 	&	0.14 	&	13.79 	\\
	&	2018-05-16	&	…	&	…	&	…	&	…	&	…	&	…	&	0.41 	&	…	\\
	&	2018-09-08	&	…	&	…	&	…	&	$^{\dagger}$1.52 	&	$^{\dagger}$40.74	&	$^{\dagger}$0.09	&	0.28 	&	5.43 	\\
	&	2018-11-03	&	…	&	…	&	…	&	$^{\dagger}$0.97 	&	$^{\dagger}$41.16 &	$^{\dagger}$0.14 	&	0.14 	&	6.93 	\\
	&	2019-01-06	&	…	&	…	&	…	&	$^{*}$0.55 	&	$^{*}$36.53 	&	$^{*}$0.06 	&	0.14 	&	3.93 	\\
	&	2019-03-03	&	39.48 	&	41.16 	&	1.69 	&	5.52 	&	40.32 	&	0.63 	&	0.28 	&	19.71 	\\
	&	2019-10-04	&	39.05 	&	40.32 	&	1.26 	&	4.97 	&	39.48 	&	0.58 	&	0.28 	&	17.75 	\\
	&	2020-01-10	&	39.48 	&	40.74 	&	1.26 	&	2.35 	&	39.48 	&	0.26 	&	0.28 	&	8.39 	\\
	&	2020-03-29	&	39.48 	&	40.74 	&	1.26 	&	3.31 	&	40.32 	&	0.34 	&	0.28 	&	11.82 	\\
	&	2020-05-29	&	39.48 	&	40.32 	&	0.84 	&	4.42	&	39.90 	&	0.44 	&	0.41	&	10.78	\\
	\enddata
 \tablecomments{As a result of visually inspecting the data, we identified features that fall below 5$\sigma$ but are likely to be detected, and these are marked with an asterisk (*). In addition, signals that exceed 5$\sigma$ but are detected in only one channel and suspected to be artificial are marked with a dagger symbol ($^\dagger$). These signals are not treated as genuine signals in the paper.}
\end{deluxetable}

%% file: SiO42_43_BasicParameter.tex
\begin{deluxetable}{crrrrrrrrr} 
	\caption{Summary of SiO maser line (42.821 and 43.122 GHz) observations.}
	\label{tab: SiO42_43MHz parameters}
	
	\tablehead{Line & Date & $V_{\rm min}$	& $V_{\rm max}$	& $V_{\rm range}$ &	$F_{\rm peak}$ & $V_{\rm peak}$	& $F_{\rm int}$	& $\rm rms$	& $\rm SNR$ \\
		&	(yyyy-mm-dd)	&	($ \rm \,km\ s^{-1}$)	&	($ \rm \,km\ s^{-1}$)	&	($ \rm \,km\ s^{-1}$)	&	(Jy)	&	($ \rm \,km\ s^{-1}$)	&	($ \rm Jy \,km\ s^{-1}$ )	&	(Jy) &  }
	
	\startdata
	\multirow{11}{*}{\makecell{SiO\\42.821\,GHz}} &	2018-01-03	&	53.88	&	54.76	&	0.88	&	0.52 	&	54.32	&	0.79 	&	0.13 	&	4.00 	\\
	&	2018-03-10	&	53.45 	&	55.20 	&	1.75 	&	1.70 	&	54.76 	&	2.62 	&	0.13 	&	13.08 	\\
	&	2018-05-16	&	53.88 	&	55.20 	&	1.31 	&	4.19 	&	54.32 	&	4.59 	&	0.26 	&	16.12 	\\
	&	2018-09-08	&	…	&	…	&	…	&	…	&	…	&	…	&	0.13 	&	…	\\
	&	2018-11-03	&	53.01 	&	53.88 	&	0.87 	&	1.31 	&	53.45 	&	5.63 	&	0.26 	&	5.04 	\\
	&	2019-01-06	&	…	&	…	&	…	&	$^{\dagger}$0.79 	&	$^{\dagger}$53.01 	&	$^{\dagger}$0.79	&	0.13 	&	6.08 	\\
	&	2019-03-03	&	…	&	…	&	…	&	$^{\dagger}$0.79 	&	$^{\dagger}$54.76 	&	$^{\dagger}$1.18	&	0.13 	&	6.08 	\\
	&	2019-10-04	&	54.32 	&	55.64 	&	1.31 	&	1.70 	&	54.76 	&	5.24 	&	0.26 	&	6.54 	\\
	&	2020-01-10	&	54.76 	&	55.64 	&	0.88 	&	0.92 	&	54.76 	&	3.14 	&	0.26 	&	3.54 	\\
	&	2020-03-29	&	54.76 	&	56.07 	&	1.31 	&	1.18 	&	55.64 	&	5.76 	&	0.26 	&	4.54 	\\
	&	2020-05-29	&	53.45 	&	56.07 	&	2.63 	&	1.97	&	55.64 	&	7.73 	&	0.26	&	7.58	\\
	\hline
	\multirow{11}{*}{\makecell{SiO\\43.122\,GHz}} &	2018-01-03	&	…	&	…	&	…	&	…	&	…	&	…	&	0.13 	&	…	\\
	&	2018-03-10	&	…	&	…	&	…	&	…	&	…	&	…	&	0.13 	&	…	\\
	&	2018-05-16	&	…	&	…	&	…	&	…	&	…	&	…	&	0.26 	&	…	\\
	&	2018-09-08	&	…	&	…	&	…	&	$^{*}$0.39 	&	$^{*}$53.82 	&	$^{*}$1.31	&	0.13 	&	3.00 	\\
	&	2018-11-03	&	…	&	…	&	…	&	$^{*}$0.66 	&	$^{*}$53.82 	&	$^{*}$1.70	&	0.26 	&	2.54 	\\
	&	2019-01-06	&	…	&	…	&	…	&	$^{\dagger}$0.66 	&	$^{\dagger}$52.52 	&	$^{\dagger}$0.00	&	0.13 	&	5.08 	\\
	&	2019-03-03	&	…	&	…	&	…	&	$^{\dagger}$0.92	&	$^{\dagger}$52.52	&	$^{\dagger}$2.10	&	0.13 	&	7.08 	\\
	&	2019-10-04	&	…	&	…	&	…	&	$^{*}$1.18 	&	$^{*}$53.38 	&	$^{*}$3.93	&	0.26 	&	4.54 	\\
	&	2020-01-10	&	…	&	…	&	…	&	$^{\dagger}$0.66 	&	$^{\dagger}$53.38 	&	$^{\dagger}$2.10	&	0.13 	&	5.08 	\\
	&	2020-03-29	&	52.08 	&	53.38 	&	1.30 	&	1.31	&	52.52	&	3.14	&	0.26 	&	5.04 	\\
	&	2020-05-29	&	…	&	…	&	…	&	$^{*}$0.79 	&	$^{*}$53.38 	&	$^{*}$1.57	&	0.26 	&	3.04 	\\
	\enddata
 \tablecomments{As a result of visually inspecting the data, we identified features that fall below 5$\sigma$ but are likely to be detected, and these are marked with an asterisk (*). In addition, signals that exceed 5$\sigma$ but are detected in only one channel and suspected to be artificial are marked with a dagger symbol ($^\dagger$). These signals are not treated as genuine signals in the paper.}
\end{deluxetable}

%% file: SiO86p24_BasicParameter.tex
\begin{deluxetable}{crrrrrr} 
	\caption{Summary of SiO maser line (86.243 GHz) observations.}
	\label{tab: SiO86.24GHz parameters}
	
	\tablehead{Line	&	Date	&	$F_{\rm peak}$ 	&	$V_{\rm peak}$	&	$F_{\rm int}$	&	$\rm rms$	&	$\rm SNR$	\\						
		&	(yyyy-mm-dd)	&	(Jy)	&	($ \rm \,km\ s^{-1}$)	&	($ \rm Jy \,km\ s^{-1}$ )	&	(Jy)	& }
	
	\startdata
	\multirow{11}{*}{\makecell{SiO\\86.243\,GHz}}&	2018-01-03	&	…	&	…	&	…	&	0.16 	&	…	\\			
	&	2018-03-10	&	$^{*}$0.64 	&	$^{*}$50.72 	&	$^{*}$1.59 	&	0.16 	&	4.00	\\
	&	2018-05-16	&	 …	&	…	&	…	&	0.48 	&	…	\\
	&	2018-09-08	&	$^{*}$0.80 	&	$^{*}$30.30 	&	$^{*}$0.95 	&	0.32 	&	2.50	\\
	&	2018-11-03	&	$^{*}$0.80 	&	$^{*}$45.94 	&	$^{*}$3.98	&	0.17	&	4.71	\\
	&	2019-01-06	&	 …	&	…	&	 …	&	0.32 	&	…	\\
	&	2019-03-03	&	…	&	…	&	…	&	0.16 	&	…	\\
	&	2019-10-04	&	…	&	…	&	…	&	0.48 	&	…	\\
	&	2020-01-10	&	$^{*}$1.11 	&	$^{*}$51.59 	&	$^{*}$5.41 	&	0.32 	&	3.47 	\\
	&	2020-03-29	&	…	&	…	&	…	&	0.32 	&	…	\\
	&	2020-05-29	&	$^{*}$1.11 	&	$^{*}$52.02 	&	$^{*}$1.43 	&	0.32	&	3.47 	\\		
	\enddata
 \tablecomments{As a result of visually inspecting the data, we identified features that fall below 5$\sigma$ but are likely to be detected, and these are marked with an asterisk (*).}
\end{deluxetable}

%% file: SiO129p36_BasicParameter.tex
\begin{deluxetable}{cll} 
	\caption{Summary of SiO maser line (129.363 GHz) observations.}
	\label{tab: SiO129.36GHz parameters}
	
	\tablehead{Line	&	Date	&	$\rm rms$	\\						
		&	(yyyy-mm-dd)	& (Jy) }	
	\startdata
	\multirow{11}{*}{\makecell{SiO\\129.363\,GHz}}&	2018-01-03	&	0.23 	\\
	&	2018-03-10	&	0.23 	\\
	&	2018-05-16	&	1.14 	\\
	&	2018-09-08	 & 	0.46 	\\
	&	2018-11-03	&	0.23 	\\
	&	2019-01-06	&	0.46 	\\
	&	2019-03-03 	&	0.46 	\\
	&	2019-10-04	&	0.91 	\\
	&	2020-01-10	& 	0.46 	\\
	&	2020-03-29	 & 	0.46 	\\
	&	2020-05-29	 & 	0.68	\\
	\enddata
\end{deluxetable}

%% file: SED_Data_Param_SingleColum.tex
\clearpage
\startlongtable
\begin{deluxetable}{ c c c c c c } 
\tablecaption{Photometric data of IRAS 19312+1950.}
\tabletypesize{\small}
\label{tab: SED of I19321}
\tablehead{
SED Filter & Wavelengh & Flux& Flux Error & R.A.(J2000)  & Dec.(J2000) \\
& $\rm (\micron)$ &$\rm(W\,m^{-2})$& $\rm(W\,m^{-2})$ &(hh mm ss.ss)&(dd mm ss.ss)}
\startdata
Johnson:V	&	0.55 	&	1.34 	$\times$ 	10$^{	-16	}$	&	0.09 	$\times$ 	10$^{	-16	}$	&	19:33:24.28	&	19:56:54.84	\\
INT/WFC:r	&	0.62 	&	3.06 	$\times$ 	10$^{	-16	}$	&	0.16 	$\times$ 	10$^{	-16	}$	&	19:33:24.28	&	19:56:54.85	\\
INT/WFC:Ha	&	0.66 	&	3.41 	$\times$ 	10$^{	-16	}$	&	0.38 	$\times$ 	10$^{	-16	}$	&	19:33:24.28	&	19:56:54.85	\\
Gaia:G	&	0.67 	&	2.82 	$\times$ 	10$^{	-16	}$	&	0.01 	$\times$ 	10$^{	-16	}$	&	19:33:24.28	&	19:56:54.84	\\
INT/WFC:i	&	0.77 	&	5.59 	$\times$ 	10$^{	-16	}$	&	0.20 	$\times$ 	10$^{	-16	}$	&	19:33:24.28	&	19:56:54.85	\\
2MASS:J	&	1.24 	&	1.13 	$\times$ 	10$^{	-13	}$	&	0.02 	$\times$ 	10$^{	-13	}$	&	19:33:24.30	&	19:56:55.00	\\
UKIDSS:J	&	1.25 	&	1.05 	$\times$ 	10$^{	-13	}$	&	0.01 	$\times$ 	10$^{	-13	}$	&	19:33:24.25	&	19:56:55.67	\\
Johnson:H	&	1.63 	&	1.56 	$\times$ 	10$^{	-12	}$	&	0.02 	$\times$ 	10$^{	-12	}$	&	19:33:24.28	&	19:56:54.84	\\
2MASS:H	&	1.65 	&	1.57 	$\times$ 	10$^{	-12	}$	&	0.03 	$\times$ 	10$^{	-12	}$	&	19:33:24.30	&	19:56:55.00	\\
2MASS:Ks	&	2.16 	&	2.13 	$\times$ 	10$^{	-12	}$	&	0.04 	$\times$ 	10$^{	-12	}$	&	19:33:24.30	&	19:56:55.00	\\
Johnson:K	&	2.19 	&	2.04 	$\times$ 	10$^{	-12	}$	&	0.04 	$\times$ 	10$^{	-12	}$	&	19:33:24.25	&	19:56:55.70	\\
WISE:W1	&	3.35 	&	4.63 	$\times$ 	10$^{	-12	}$	&	0.38 	$\times$ 	10$^{	-12	}$	&	19:33:24.24	&	19:56:55.50	\\
WISE:W2	&	4.60 	&	1.30 	$\times$ 	10$^{	-11	}$	&	0.05 	$\times$ 	10$^{	-11	}$	&	19:33:24.29	&	19:56:56.02	\\
AKARI:S9W	&	8.61 	&	4.77 	$\times$ 	10$^{	-12	}$	&	0.03 	$\times$ 	10$^{	-12	}$	&	19:33:24.23	&	19:56:55.14	\\
WISE:W3	&	11.57 	&	5.72 	$\times$ 	10$^{	-12	}$	&	0.09 	$\times$ 	10$^{	-12	}$	&	19:33:24.29	&	19:56:56.02	\\
MSX:C	&	12.14 	&	5.61 	$\times$ 	10$^{	-12	}$	&	0.27 	$\times$ 	10$^{	-12	}$	&	19:33:24.17	&	19:56:55.32	\\
MSX:D	&	14.62 	&	4.74 	$\times$ 	10$^{	-12	}$	&	0.29 	$\times$ 	10$^{	-12	}$	&	19:33:24.17	&	19:56:55.32	\\
AKARI:L18W	&	18.39 	&	6.03 	$\times$ 	10$^{	-12	}$	&	0.08 	$\times$ 	10$^{	-12	}$	&	19:33:24.23	&	19:56:55.14	\\
MSX:E	&	21.41 	&	6.37 	$\times$ 	10$^{	-12	}$	&	0.39 	$\times$ 	10$^{	-12	}$	&	19:33:24.17	&	19:56:55.32	\\
WISE:W4	&	22.04 	&	7.81 	$\times$ 	10$^{	-12	}$	&	0.02 	$\times$ 	10$^{	-12	}$	&	19:33:24.29	&	19:56:56.02	\\
IRAS:25$\mu$m &	23.79 	&	8.90 	$\times$ 	10$^{	-12	}$	&	0.35 	$\times$ 	10$^{	-12	}$	&	19:33:24.35	&	19:56:54.81	\\
IRAS:60$\mu$m &	61.81 	&	1.78 	$\times$ 	10$^{	-11	}$	&	0.23 	$\times$ 	10$^{	-11	}$	&	19:33:24.35	&	19:56:54.81	\\
PACS:70$\mu$m &	70.04 	&	2.31 	$\times$ 	10$^{	-11	}$	&	0.08 	$\times$ 	10$^{	-11	}$	&	19:33:24.21	&	19:56:57.24	\\
IRAS:100$\mu$m &	101.97 	&	1.24 	$\times$ 	10$^{	-11	}$	&	0.30 	$\times$ 	10$^{	-11	}$	&	19:33:24.35	&	19:56:54.81	\\
PACS:160$\mu$m &	160.32 	&	4.53 	$\times$ 	10$^{	-12	}$	&	0.09 	$\times$ 	10$^{	-12	}$	&	19:33:24.21	&	19:56:57.24	\\
SPIRE:250$\mu$m &	249.83 	&	1.93 	$\times$ 	10$^{	-12	}$	&	0.05 	$\times$ 	10$^{	-12	}$	&	19:33:24.21	&	19:56:57.24	\\
SPIRE:350$\mu$m &	350.00 	&	4.13 	$\times$ 	10$^{	-13	}$	&	0.06 	$\times$ 	10$^{	-13	}$	&	19:33:24.21	&	19:56:57.24	\\
SPIRE:500$\mu$m &	500.00 	&	1.34 	$\times$ 	10$^{	-13	}$	&	0.03 	$\times$ 	10$^{	-13	}$	&	19:33:24.21	&	19:56:57.24	\\
ATLASGAL:870$\mu$m &	870.00 	&	3.86 	$\times$ 	10$^{	-15	}$	&	0.44 	$\times$ 	10$^{	-15	}$	&	19:33:24.21	&	19:56:57.24	\\
\enddata
\end{deluxetable}